\documentclass[12pt]{article}
\usepackage{a4wide,epsfig}
\usepackage[sort]{cite}

\begin{document}
\newcommand{\be}{\begin{equation}}
\newcommand{\ee}{\end{equation}}
\newcommand{\bfm}[1]{\mbox{\boldmath$#1$}}
\newcommand{\bff}[1]{\mbox{\scriptsize\boldmath${#1}$}}

\newcommand{\al}{\alpha}
\newcommand{\bt}{\beta}
\newcommand{\lm}{\lambda}
\newcommand{\bea}{\begin{eqnarray}}
\newcommand{\eea}{\end{eqnarray}}
\newcommand{\gm}{\gamma}
\newcommand{\Gm}{\Gamma}
\newcommand{\dl}{\delta}
\newcommand{\Dl}{\Delta}
\newcommand{\ep}{\epsilon}
\newcommand{\kp}{\kappa}
\newcommand{\Lm}{\Lambda}
\newcommand{\om}{\omega}
\newcommand{\pa}{\partial}
\newcommand{\nn}{\nonumber}
\newcommand{\dd}{\mbox{d}}
\newcommand{\uk}{\underline{k}}
\newcommand{\gsim}{\;\rlap{\lower 3.5 pt \hbox{$\mathchar \sim$}} \raise 1pt \hbox {$>$}\;}
\newcommand{\lsim}{\;\rlap{\lower 3.5 pt \hbox{$\mathchar \sim$}} \raise 1pt \hbox {$<$}\;}

\title{
\begin{flushleft}
\normalsize PSI-PR-05-04 \\
TTP05-17 \\
hep-ph/0509157 \\
\end{flushleft}
\vspace{1.0cm}
Two-Loop Electroweak Logarithms in Four-Fermion \\[1mm]
Processes at High Energy}
\author{
  {\large B.~Jantzen}\footnote{B. Feucht in previous publications.}~$^{a,b}$,\,
  {\large J.H.~K\"uhn} $^b$,\,
  {\large A.A.~Penin} $^{b,c}$,\, and
  {\large V.A.~Smirnov} $^{d,e}$\\[3mm]
  $^a${\small {\it Paul Scherrer Institut}}\\
  {\small {\it 5232 Villigen PSI, Switzerland}}\\[2mm]
  $^b${\small {\it Institut f{\"u}r Theoretische Teilchenphysik,
  Universit{\"a}t Karlsruhe}}\\
  {\small {\it 76128 Karlsruhe, Germany}}\\[2mm]
  $^c${\small {\it Institute for Nuclear
  Research of Russian Academy   of Sciences}}\\
  {\small {\it 117312 Moscow, Russia}}\\[2mm]
  $^d${\small {\it  Nuclear Physics
  Institute of Moscow State University}}\\
  {\small {\it  119992 Moscow, Russia}}\\[2mm]
  $^e${\small {\it II. Institut f{\"u}r Theoretische Physik,
  Universit{\"a}t Hamburg}}\\
  {\small {\it  22761 Hamburg, Germany}}
}
\date{}
\maketitle
\begin{abstract}
  We present the complete analytical result for the two-loop
  logarithmically enhanced contributions to the high energy asymptotic
  behavior of the vector form factor and the four-fermion cross section
  in a spontaneously broken $SU(2)$ gauge model. On the basis of this
  result we derive the dominant two-loop electroweak corrections to the
  neutral current four-fermion processes at high energies.  Previously
  neglected effects of the gauge boson mass difference are included
  through the next-to-next-to-leading logarithmic approximation.
  \\[2mm]
\end{abstract}


\section{Introduction}
\label{intr}
Recently a new wave of interest in the Sudakov asymptotic regime
\cite{Sud,Jac} has risen in connection with higher-order corrections to
electroweak processes at high energies
\cite{Kur,DegSir,Bec,CiaCom,KuhPen,Fad,KPS,KMPS,Bec2,BRV,DenPoz,HKK,BeeWet,FKM,
DMP,Bec3,Bec4,Poz,FKPS}.
Experimental and theoretical studies of electroweak interactions have
traditionally explored the range from very low energies, {\it e.g.}
through parity violation in atoms, up to energies comparable to the
masses of the $W$- and $Z$-bosons at LEP or the Tevatron.  The advent of
multi-TeV colliders like the LHC during the present decade or a future
linear electron-positron collider will give access to a completely new
energy domain. Once the characteristic energies $\sqrt{s}$ are far
larger than the masses of the $W$- and $Z$-bosons, $M_{W,Z}$, exclusive
reactions like electron-positron (or quark-antiquark) annihilation into
a pair of fermions or gauge bosons will receive virtual corrections
enhanced by powers of the large {\it electroweak} logarithm
$\ln\bigl({s/ M_{W,Z}^2}\bigr)$.

In Refs.~\cite{KPS,KMPS} we have extended the leading logarithmic (LL)
analysis of \cite{Fad}. The next-to-leading logarithmic (NLL) and
next-to-next-to-leading logarithmic (NNLL) corrections to the high
energy asymptotic behavior of the neutral current four-fermion processes
have been resummed to all orders using the {\it evolution equation}
approach. Only the light quark case was considered and the mass
difference between the neutral and charged gauge bosons was neglected.
On the basis of this result the logarithmically enhanced part of the
phenomenologically important two-loop corrections to the total cross
section and to various asymmetries was obtained including the
$\ln^n(s/M_{W,Z}^2)$ terms with $n=2,~3,~4$.  The results up to NLL have
been confirmed by the explicit one-loop \cite{Bec,BRV,DenPoz} and
two-loop \cite{HKK,BeeWet,FKM,DMP,Poz} calculations.  The subleading
logarithms in the TeV region are comparable to the leading terms due to
their large numerical coefficients.  Thus, the calculation of the
remaining two-loop linear logarithms is necessary to control the
convergence of the logarithmic expansion.  These corrections represent
the next-to-next-to-next-to-leading logarithmic (N$^3$LL) contribution.
They are of special interest both from the phenomenological and
conceptual point of view because, in contrast to the higher powers of
the logarithm, they are sensitive to the details of the gauge boson mass
generation.  The first results beyond the NNLL approximation have been
obtained in \cite{FKM,FKPS,JKPS}.

In this paper we extend our previous analysis and complete the
calculation of the two-loop logarithmic corrections to the neutral
current four-fermion processes.  We incorporate previously neglected
effects of the gauge boson mass difference.  The two-loop logarithmic
terms are derived within the {\it expansion by regions} approach
\cite{BenSmi,SmiRak,Smi1,Smi2} by inspecting the structure of
singularities of the contributions of different regions.  The
calculation is significantly simplified by taking the exponentiation of
the logarithmic corrections in the Sudakov limit into account.  This
property naturally appears and can be fully elaborated within the
evolution equation approach \cite{Mue1,Col,Sen1}.  To identify the pure
QED infrared logarithms which are compensated by soft real photon
radiation we combine the {\it hard} evolution equation which governs the
dependence of the amplitudes on $s$ with the {\it infrared} evolution
equation \cite{Fad} which describes the dependence of the amplitude on
an infrared regulator. The main results of the present paper have been
reported earlier in a short letter \cite{JKPS}.

In Section~\ref{secf} we present the complete result for the two-loop
logarithmic corrections to the form factor which describes fermion
scattering in an external Abelian field. Two models are considered: a
non-Abelian $SU(2)$ gauge theory with the Higgs mechanism of gauge boson
mass generation and an Abelian $U(1)\times U(1)$ theory with two gauge
bosons of essentially different masses \cite{FKPS}.  In
Section~\ref{seca} we generalize the result to the four-fermion process.
Finally, in Section~\ref{secew}, we apply this result to electroweak
processes.  A brief summary and conclusions are given in
Section~\ref{summ}.

\section{Abelian form factor in the Sudakov limit}
\label{secf}
The vector form factor ${\cal F}$ determines the fermion scattering
amplitude in an external Abelian field.  It plays a special role since
it is the simplest quantity which includes the complete information
about the universal {\it collinear} logarithms.  This information is
directly applicable to a process with an arbitrary number of fermions.
The form factor is a function of the Euclidean momentum transfer
$Q^2=-(p_1-p_2)^2$ where $p_{1,2}$ is the incoming/outgoing fermion
momentum.  In the next two sections we consider two characteristic
examples: (i) the $SU(2)$ gauge model with gauge bosons of a nonzero
mass $M$ which emulates the massive gauge boson sector of the standard
model and (ii) the $U(1)\times U(1)$ gauge model with two gauge bosons
of essentially different masses which emulates the effect of the $Z-\gm$
mass gap.  We focus on the asymptotic behavior of the form factor in the
Sudakov limit $M/Q\ll 1$ with on-shell massless fermions,
$p_1^2=p_2^2=0$.

In the Sudakov limit the coefficients of the perturbative series in the
coupling constant can be expanded in $M^2/Q^2$.  To compute the leading
term of the series in $M^2/Q^2$ we use the expansion by regions approach
\cite{BenSmi,SmiRak,Smi1,Smi2}.  It is based on separating the
contributions of dynamical modes or {\it regions} characteristic for
different asymptotic regimes and consists of the following steps:
\begin{itemize}
\item[(i)] consider various regions of a loop momentum $k$ and expand,
  in every region, the integrand in a Taylor series with respect to the
  parameters considered small in this region;
\item[(ii)] integrate the expanded integrand over the whole integration
  domain of the loop momenta;
\item[(iii)] put to zero any scaleless integral.
\end{itemize}
In step (ii) dimensional regularization with $d=4-2\ep$ space-time
dimensions is used to handle the divergences.  In the Sudakov limit
under consideration the following regions are relevant
\cite{Ste1,LibSte,Mue2}:
\bea
\mbox{{\it hard} (h):} && k\sim Q\,,
\nn \\
\mbox{{\it 1-collinear} (1c):} && k_+\sim Q,\,\,k_-\sim M^2/Q\,,
\,\, \uk \sim M\,,
\nn \\
\mbox{{\it 2-collinear} (2c):} && k_-\sim Q,\,\,k_+\sim M^2/Q\,,
\,\,\uk \sim M \,,
\nn \\
\mbox{{\it soft} (s):} && k\sim M\,,
\label{reg}
\eea
where $k_{\pm} =k_0\pm k_3, \, \uk=(k_1,k_2)$ and we choose $p_{1,2} =
(Q/2,0,0,\mp Q/2)$ so that $2 p_1 p_2 = Q^2=-s$.  By $k\sim Q$, {\it
  etc.} we mean that every component of $k$ is of order $Q$.  The
expansion procedure can be facilitated by a technique based on
Mellin-Barnes representation which can be used not only for evaluating
Feynman integrals but also to pick up contributions of regions (see
Sect. 4.8 of \cite{Smi3} for a general discussion and \cite{Smi4} where
the MB representation was used for this purpose.)

\subsection{$SU(2)$ model with massive gauge boson}
\label{secsym}
Let us apply the method to compute the corrections to the form factor in
the $SU(2)$ model.  In one loop the expansion by regions leads to the
following decomposition
\be
{\cal F}^{(1)}=
{\cal F}^{(1)}_h+{\cal F}^{(1)}_c+{\cal F}^{(1)}_s\,,
\label{1loopfreg}
\ee
where the subscript $c$ denotes the contribution of both collinear
regions, for a perturbative function $f(\al)$ we define
\be
f(\al)=\sum_n\left(\al\over 4\pi\right)^nf^{(n)}\,,
\label{defexp}
\ee
and the form factor in the Born approximation is normalized to 1.  The
hard region contribution, which we will later need, reads
\bea
{\cal F}^{(1)}_h&=&{C_F\over Q^{2\ep}}\left[-{2\over\ep^2}-{3\over\ep}
+{\pi^2\over 6}-8+\left(-16+{\pi^2\over4}+{14\over
    3}\zeta(3)\right)\ep\right]
+{\cal O}(\ep^2)
\,,
\label{1loopfh}
\eea
where $C_F=(N^2-1)/(2N)$ for a $SU(N)$ gauge group,
$\zeta(3)=1.202057\ldots$ is the value of the Riemann's zeta-function
and all the power-suppressed terms are neglected. For convenience we do
not include the standard factor $(4\pi e^{-\gm_E\ep}(\mu^2))^{\ep}$ per
loop, where $\gm_E=0.577216\ldots$ is Euler's constant.  The
contributions of all the regions \cite{KMPS} add up to the well known
finite result
\be
{\cal F}^{(1)}=-C_F\left({\cal L}^2
-3{\cal L}+{7\over 2}+{2\pi^2\over 3}\right)\,,
\label{1loopfsum}
\ee
where ${\cal L}=\ln\left({Q^2/M^2}\right)$.  A similar decomposition can
be performed in two loops
\be
{\cal F}^{(2)}={\cal F}^{(2)}_{hh}+{\cal F}^{(2)}_{hc}
+{\cal F}^{(2)}_{cc}+\ldots
\label{2loopfreg}
\ee
The hard-hard part  reads
\bea
{\cal F}^{(2)}_{hh}&=&
\left({1\over 2}{\cal F}^{(1)}_{h}-{\beta_0\over\ep}\right)
{\cal F}^{(1)}_{h}+{C_F\over Q^{4\ep}}\left\{
\left[-{11\over 6\ep^3}
+\left(-{83\over 9}+{\pi^2\over 6}\right){1\over \ep^2}
+\left(-{4129\over 108}-{11\over 36}\pi^2\right.\right.\right.
\nn \\
&&
\left.+13\zeta(3)\bigg){1\over\ep}\right]
C_A+\left[{2\over 3\ep^3}+{28\over 9\ep^2}+
\left({353\over 27}+{\pi^2\over 9}\right){1\over \ep}\right]T_Fn_f
+\left[{1\over 6\ep^3}+{17\over 18\ep^2}\right.
\nn \\
&&
\left.\left.
+\left({455\over 108}+{\pi^2\over 36}\right){1\over\ep}\right]T_Fn_s
+\left[-{3\over 4}+\pi^2-12\zeta(3)\right]{C_F\over\ep}
\right\}+{\cal O}(\ep^0)\,,
\label{2loopfhh}
\eea
for $\al$ defined in the $\overline{MS}$ scheme.  Here $C_A=N$,
$T_F=1/2$, $\bt_0=11C_A/3-4T_Fn_f/3-T_Fn_s/3$ is the one-loop
beta-function, and $n_f$ ($n_s$) is the number of Dirac fermions
(scalars) in the fundamental representation.  Eq.~(\ref{2loopfhh})
coincides with the known massless two-loop QCD result \cite{KraLam,MMN}
except for the scalar loop contribution which is new.  With $\al$
renormalized at the scale $M$ in the one-loop result, the total two-loop
contribution takes the form
\bea
{\cal F}^{(2)}&=&{1\over 2}{{\cal F}^{(1)}}^2+C_F\left\{
\left[{11\over 9}{\cal L}+\left(-{233\over 18}+{\pi^2\over 3}\right)
\right]C_A+\left[-{4\over 9}{\cal L}+{38\over 9}
\right]T_Fn_f+\left[-{1\over 9}{\cal L}+{25\over 18}
\right]\right.
\nn \\
&&
\times T_Fn_s\Bigg\}{\cal L}^2
+\left[\Delta^{(2)}_{NA}+\Delta^{(2)}_f+\Delta^{(2)}_s+\Delta^{(2)}_A
\right]{\cal L}+{\cal O}({\cal L}^0) \,.
\label{2looplog}
\eea
The coefficients of the second and higher powers of the logarithm are
insensitive to the infrared structure of the model as explained below.
In contrast, the coefficient of the linear logarithm does in general
depend on the whole mass spectrum of the model.  The purely Abelian term
reads \cite{FKPS}
\bea
\Delta^{(2)}_A&=&\left({3\over 2}-2\pi^2+24\zeta(3)\right)C_F^2\,.
\label{2loopab}
\eea
Massless Dirac fermions give \cite{FKM}
\bea
\Delta^{(2)}_f&=&-{34\over 3}C_FT_Fn_f\,.
\label{2loopf}
\eea
The non-Abelian contribution depends on the details of the gauge boson
mass generation.  For the spontaneously broken gauge group $SU(2)$ with
a single Higgs boson of mass $M_H=M$ in the fundamental representation
explicit calculation leads to
\bea
\Delta^{(2)}_{NA}+\Delta^{(2)}_s&=&
{749\over 16}+{43\over 24}\pi^2
-44\zeta(3)+{15\over 4}\sqrt{3}\pi
+{13\over 2}\sqrt{3}{\rm Cl}_2\left({\pi\over 3}\right)\,.
\label{2loopna}
\eea
Here ${\rm Cl}_2\left({\pi\over 3}\right)=1.014942\ldots$ is the value
of the Clausen function.  For comparison, in the (hypothetical) case of
a light Higgs boson $M_H\ll M$ this contribution becomes
\bea
\Delta^{(2)}_{NA}+\Delta^{(2)}_s&=&
{747\over 16}+{97\over 48}\pi^2
-44\zeta(3)+{33\over 8}\sqrt{3}\pi
+{21\over 4}\sqrt{3}{\rm  Cl}_2\left({\pi\over 3}\right)\,.
\label{2loopna1}
\eea
In the electroweak theory inspired model with the $SU(2)_L$ gauge group,
six left-handed fermion doublets, and  $M_H=M$, the result for
the two-loop logarithmic corrections reads
\bea
{\cal F}^{(2)}&=&{9\over 32}{\cal L}^4
-{19\over 48}{\cal L}^3-\left({463\over 48}-{7\over 8}\pi^2\right){\cal L}^2
+\left(29-{11\over 24}\pi^2-{61\over 2}\zeta(3)+{15\over 4}\sqrt{3}\pi
\right.
\nn\\
&&
+{13\over 2}\sqrt{3}{\rm Cl}_2\left({\pi\over 3}\right)\bigg)
{\cal L}+{\cal O}({\cal L}^0)\,.
\label{fsu2}
\eea
The asymptotic dependence of the form factor on $Q$ in the Sudakov limit is
governed by the hard evolution equation \cite{Mue1,Col,Sen1}
\be
{\partial\over\partial\ln{Q^2}}{\cal F}=
\left[\int_{M^2}^{Q^2}{\dd x\over x}\gm(\al(x))+\zeta(\al(Q^2))
+\xi(\al(M^2)) \right] {\cal F} \,.
\label{evoleqf}
\ee
Its solution is
\be
{\cal F}=F_0(\al(M^2))\exp \left\{\int_{M^2}^{Q^2}{\dd x\over x}
\left[\int_{M^2}^{x}{\dd x'\over x'}\gm(\al(x'))+\zeta(\al(x))
+\xi(\al(M^2))\right]\right\} \,.
\label{evolsolf}
\ee
By calculating the functions entering the evolution equation order by
order in $\al$ one gets the logarithmic approximations for the form
factor. For example, the LL approximation includes all the terms of the
form $\al^n{\cal L}^{2n}$ and is determined by the one-loop value of
$\gm(\al)$; the NLL approximation includes all the terms of the form
$\al^n{\cal L}^{2n-m}$ with $m=0,~1$ and requires the one-loop values of
$\gm(\al)$, $\zeta(\al)$ and $\xi(\al)$ as well as the one-loop running
of $\al$ in $\gm(\al)$; and so on.  The functions entering the evolution
equation can in principle be determined by comparing
Eq.~(\ref{evolsolf}) expanded in the coupling constant to the fixed
order result for the form factor.  Within the expansion by regions
approach the logarithmic contributions show up as singularities of the
different regions.  One can identify the regions relevant for
determining a given parameter of the evolution equation and compute them
separately up to the required accuracy which facilitates the analysis by
far.  For example, the anomalous dimensions $\gm(\alpha)$ and
$\zeta(\alpha)$ are known to be mass-independent and determined by the
singularities of the contribution with all the loop momenta being hard
\cite{Mue1,Col,Sen1}. The dimensionally regularized hard contribution
exponentiates as well \cite{Sen1,MagSte} with the functions
$\gm(\alpha)$ and $\zeta(\alpha)$ parameterizing the double and single
pole contribution to the exponent. The one- and two-loop hard
contribution can be written as
\bea
{\cal F}^{(1)}_h&=&
{1\over Q^{2\ep}}\left({\gm^{(1)}\over\ep^2}-
{\zeta^{(1)}\over\ep} +F_0{}_h^{(1)}\right) +{\cal O}(\ep) \,, \nn
\\
{\cal F}^{(2)}_{hh}&=&\left({1\over 2}{\cal F}^{(1)}_{h}-{\beta_0\over \ep}\right)
{\cal F}^{(1)}_{h}+{1\over Q^{4\ep}}\left\{
\left[{1\over \ep^3}{\gm^{(1)}\beta_0\over 4}
+{1\over \ep^2}\left({\gm^{(2)}\over 4}-{\zeta^{(1)}\bt_0\over 2}\right)
\right.\right.
\nn\\
&&
+{1\over 2\ep}\left(-\zeta^{(2)}+F_0{}_h^{(1)}\bt_0\right)\Bigg]\Bigg\}
+{\cal O}(\ep^0)\,,
\label{2loopfhhdec}
\eea
From Eqs.~(\ref{1loopfh},~\ref{2loopfhh},~\ref{2loopfhhdec})  we find
\bea
\gm^{(1)}&=&-2C_F \,,
\nn \\
\zeta^{(1)}&=&3C_F\,,
\nn
\\
\gm^{(2)}&=&C_F\left[\left(-{134\over 9}+{2\over 3}\pi^2\right)C_A
+{40\over 9}T_Fn_f+{16\over 9}T_Fn_s\right] \,,
\nn
\\
\zeta^{(2)}&=&C_F\left[\left({2545\over 54}+{11\over 9}\pi^2
-{26}\zeta(3)\right)C_A
\right.-\left({418\over 27}+{4\over 9}\pi^2\right)T_Fn_f
\nn\\
&&
\left.-\left({311\over 54}+{\pi^2\over 9}\right)T_Fn_s+\left({3\over 2}-{2}\pi^2
+{24}\zeta(3)\right)C_F\right] \,.
\label{2loopz}
\eea
At the same time the functions $\xi(\al)$ and $F_0(\al)$ fix the initial
conditions for the evolution equation at $Q=M$ and do depend on the
infrared structure of the model.  To determine the function $\xi(\al)$
one has to know the singularities of the collinear region contribution
while $F_0(\al)$ requires the complete information on the contributions
of all the regions.  The total one- and two-loop form factor can be
expressed through the parameters of the evolution equation as follows
\bea
{\cal F}^{(1)}&=&{1\over 2}\gm^{(1)}{\cal L}^2
+\left(\xi^{(1)}+\zeta^{(1)}\right){\cal L}
+F_0^{(1)}\,.
\nn
\\
{\cal F}^{(2)}&=&
{1\over 8}{(\gm^{(1)})^2}{\cal L}^4
+{1\over 2}\left(\xi^{(1)}+\zeta^{(1)}-{1\over 3}\bt_0\right)
\gm^{(1)}{\cal L}^3+{1\over 2}\left(\gm^{(2)}+
\left(\xi^{(1)}+\zeta^{(1)}\right)^2\right.
\nn \\
&&
-\bt_0\zeta^{(1)}+F_0^{(1)}\gm^{(1)}\bigg){\cal L}^2
+\left({\zeta^{(2)}+\xi^{(2)}+F_0^{(1)}(\zeta^{(1)}+\xi^{(1)})}\right)
{\cal L}+{\cal O}({\cal L}^0)\,.
\label{2loopfdec}
\eea
With the known values of $\gm^{(1,2)}$ and $\zeta^{(1,2)}$ it is
straightforward to obtain the result for the remaining functions
\bea
\xi^{(1)} &=&0 \,,
\nn\\
F_0^{(1)}&=& -C_F
\left({7\over 2}+{2\pi^2\over 3}\right) \,
\nn
\\
\xi^{(2)}&=&\xi^{(2)}_{NA}+\xi^{(2)}_{f}+\xi^{(2)}_{s}+\xi^{(2)}_{A}\,,
\label{2loopx}
\eea
where the Abelian contribution vanishes $\xi^{(2)}_{A}=0$, the massless
Dirac fermions give
\be
\xi^{(2)}_{f}=\left({112\over 27}
+{4\over 9}\pi^2\right)C_FT_Fn_f\,,
\label{2loopxf}
\ee
and for the spontaneously broken $SU(2)$ model with $M_H=M$ we get
\be
\xi^{(2)}_{NA}+\xi^{(2)}_{s}=-{391\over 18}
-5\zeta(3)+{15\over 4}\sqrt{3}\pi
+{13\over 2}\sqrt{3}{\rm Cl}_2\left({\pi\over 3}\right)\,.
\label{2loopxnas}
\ee
Note that the functions $\gm(\al)$ and $\xi(\al)$ are protected against
the Abelian multiloop corrections by the properties of the light-cone
Wilson loop \cite{Col,Sen1,KorRad}.

The analysis of the evolution equation gives a lot of insight into the
structure of the logarithmic corrections. For example,
Eq.~(\ref{2loopfdec}) tells us that, up to the NNLL approximation, the
information on the infrared structure of the model enters through the
one-loop coefficients $\xi^{(1)}$ and $F_0^{(1)}$ which are insensitive
to the details of the mass generation.  Thus one can compute the
coefficient of the two-loop quadratic logarithm with the gauge boson
mass introduced by hand \cite{KMPS}.

Note that to complete the N$^3$LL approximation one needs the three-loop
value of $\gm(\al)$. For $n_s=0$ this has been recently obtained in the
context of QCD splitting functions \cite{MVV}
\bea
\gm^{(3)}&=&C_F\left[\left(-{245\over 3}+{268\over 27}\pi^2
-{44\over 3}\zeta(3)-{22\over 45}\pi^4\right)C_A^2
+\left({836\over 27}-{80\over 27}\pi^2
+{112\over 3}\zeta(3)\right)C_AT_Fn_f\right.
\nn\\
&&\left.+\left({110\over 3}-{32}\zeta(3)\right)C_FT_Fn_f+
{32\over 27}(T_Fn_f)^2\right]\,.
\label{3loopgam}
\eea

\subsection{$U(1)\times U(1)$ model with mass gap}
\label{secgap}
Let us now discuss the second example, a $U(1)\times U(1)$
model with  $\lm$, $\al'$  and $M$, $\al$ for masses and coupling
constants, respectively.  We consider the limit
$\lm\ll M$ and make use of  the {\it infrared} evolution equation
which governs the dependence of the form factor ${\cal F}(\lm,M,Q)$ on
$\lm$ \cite{Fad}.  The virtual corrections become divergent in the limit
$\lm\to 0$.  According to the Kinoshita-Lee-Nauenberg theorem
\cite{Kin,LeeNau}, these divergences are cancelled against the ones of
the corrections due to the emission of real light gauge bosons of
vanishing energy and/or collinear to one of the on-shell fermion lines.
The singular behavior of the form factor must be the same in the full
$U_{\al'}(1)\times U_{\al}(1)$ theory and the effective $U_{\al'}(1)$
model with only the light gauge boson.  For $\lm\ll M\ll Q$ the solution
of the infrared evolution equation is given by the Abelian part of the
exponent~(\ref{evolsolf}) with $M$, $\al$ replaced by $\lm$, $\al'$.
Thus the form factor can be written in a factorized form
\be
{\cal F}(\lm,M,Q)=
\tilde{F}(M,Q){\cal F}_{\al'}(\lm,Q)+{\cal O}(\lm/M)\,,
\label{fac}
\ee
where ${\cal F}_{\al'}(\lm,Q)$ stands for the $U_{\al'}(1)$ form factor
and $\tilde{F}(M,Q)$ depends both on $\al$ and $\al'$, and incorporates
all the logarithms of the form $\ln\left({Q^2/M^2}\right)$.  It can be
obtained directly by calculating the ratio
\be
\tilde{F}(M,Q)=\left[{{\cal F}(\lm,M,Q)\over
{\cal F}_{\al'}(\lm,Q)}\right]_{\lm\to 0} \,.
\label{tf}
\ee
Since the function $\tilde{F}(M,Q)$ does not depend on the infrared
regularization, the ratio in Eq.~(\ref{tf}) can be evaluated with
$\lm=0$ using dimensional regularization for the infrared divergences.
The resulting two-parameter perturbative expansion is
\be
\tilde{F}(M,Q)=\sum_{n,m}{\al'{}^n\al^m\over
  (4\pi)^{n+m}}\tilde{F}^{(n,m)}
\ee
where
\be
\tilde{F}^{(0,0)}=1,\qquad \tilde{F}^{(n,0)}=0, \qquad
\tilde{F}^{(0,m)}={\cal F}^{(m)}\,,
\ee
and   the two-loop interference term reads \cite{FKPS}
\bea
\tilde{F}^{(1,1)}&=&
\left(3-4\pi^2+48\zeta(3)\right){\cal L}+{\cal O}({\cal L}^0)\,.
\label{2looptf}
\eea
In the equal mass case, $\lm=M$, we have an additional
reparameterization symmetry, and the form factor is determined by
Eq.~(\ref{evolsolf}) with the effective coupling $\bar\al= \al'+\al$ so
that ${\cal F}(M,M,Q)={\cal F}_{\bar\al}(M,Q)$.  We can now write down
the matching relation
\be
{\cal F}(M,M,Q)=
C(M,Q)\tilde{F}(M,Q){\cal F}_{\al'}(M,Q)\,,
\label{matf}
\ee
where the matching coefficient $C(M,Q)$ represents the effect of the
power-suppressed terms neglected in Eq.~(\ref{fac}).  By combining the
explicit results for ${\cal F}_{\al'}(M,Q)$ and $\tilde{F}(M,Q)$ the
matching coefficient has found to be $C(M,Q)=1+{\cal O}({\al'\al{\cal
    L}^0})$ \cite{FKPS}.  In two-loops it does not contain logarithmic
terms, and up to the N$^3$LL accuracy, the product $\tilde{F}(M,Q){\cal
  F}_{\al'}(\lm,Q)$ continuously approaches ${\cal F}(M,M,Q)$ as $\lm$
goes to $M$. Therefore, to get {\it all} the logarithms of the heavy
gauge boson mass in two-loop approximation for the theory with mass gap,
it is sufficient to divide the form factor ${\cal F}_{\bar\al}(M,Q)$ of
the symmetric phase by the form factor ${\cal F}_{\al'}(\lm,Q)$ of the
effective $U_{\al'}(1)$ theory taken at the symmetric point $\lm=M$.
Thus we have reduced the calculation in the theory with mass gap to the
one in the symmetric theory with a single mass parameter.  The
logarithmic terms in the expansion of $\tilde{F}(M,Q)$ exponentiate by
construction and one can describe the exponent with a set of functions
$\tilde\gm(\al,\al')$, $\tilde\zeta(\al,\al')$, $\tilde\xi(\al,\al')$
and $\tilde{F}_0(\al,\al')$ in analogy with Eq.~(\ref{evolsolf}).  The
matching procedure can naturally be formulated in terms of these
functions.  For the mass-independent functions we have the all order
relation
\bea
\tilde\gm(\al',\al)&=&\gm(\bar\al)-\gm(\al')\,,
\nn\\
\tilde\zeta(\al',\al)&=&\zeta(\bar\al)-\zeta(\al')\,.
\label{hardmat}
\eea
In two loops we obtain by explicit calculation
\be
\tilde\xi(\al',\al)=\xi(\bar\al)=0\,,
\label{ximat}
\ee
which holds in higher orders for the Abelian model due to the
nonrenormalization properties discussed in the previous section.  Thus,
the only nontrivial two-loop matching is for the coefficient
$\tilde{F}_0^{(1,1)}$ due to the non-logarithmic contribution to
$C(M,Q)$ which is beyond the accuracy of our analysis.

Note that the absence of the two-loop linear-logarithmic term in
$C(M,Q)$ is an exceptional feature of the Abelian corrections.  The
general analysis of the evolution equation \cite{KMPS} shows that the
terms neglected in Eq.~(\ref{fac}) contribute starting from the N$^3$LL
approximation.  Indeed, the solution of the hard evolution equation for
$F(\lm,M,Q)$ which determines its dependence on $Q$ is of the
form~(\ref{evolsolf}) with the infrared sensitive quantities $F_0$ and
$\xi$ being functions of the ratio $\lm/M$.  A nontrivial dependence on
the mass ratio in general emerges first through the two-loop coefficient
$\xi^{(2)}$ due to the interference diagrams with both massive and
massless gauge bosons.  The matching is necessary to take care of the
difference $\xi^{(2)}|_{\lm/M=1}\ne\xi^{(2)}|_{\lm/M=0}$.  Thus, for a
non-Abelian theory with the mass gap of the standard model type, {\it
  i.e.} with interaction between the heavy and light gauge bosons, the
matching becomes nontrivial already in N$^3$LL approximation.

\section{Four-fermion amplitude}
\label{seca}
We consider the four-fermion scattering at fixed angles in the limit
where all the kinematical invariants are of the same order and are far
larger than the gauge boson mass, $|s|\sim |t| \sim |u| \gg M^2$.  The
analysis of the four-fermion amplitude is complicated by the additional
kinematical variable and the presence of different isospin and Lorentz
structures.  We adopt the following notation
\bea
{\cal A}^\lm&=&
\bar{\psi_2}t^a\gm_\mu{\psi_1}
\bar{\psi_4}t^a\gm_\mu{\psi_3}\,, \nn \\
{\cal A}^\lm_{LL}&=&
\bar{\psi_2}_Lt^a\gm_\mu{\psi_1}_L
\bar{\psi_4}_Lt^a\gm_\mu{\psi_3}_L\,,
\nn\\
\label{basis}
{\cal A}_{LR}^d&=&
\bar{\psi_2}_L\gm_\mu{\psi_1}_L
\bar{\psi_4}_R\gm_\mu{\psi_3}_R\,,
\eea
{\it etc.} Here $t^a$ denotes the $SU(N)$ ``isospin'' generator, $p_i$
the momentum of the $i$th fermion and $p_1$, $p_3$ are incoming, and
$p_2$, $p_4$ outgoing momenta respectively.  Hence $t=(p_1-p_4)^2=-sx_-$
and $u=(p_1+p_3)^2=-sx_+$ where $x_{\pm}=(1\pm\cos\theta)/2$ and
$\theta$ is the angle between $\bfm{p}_1$ and $\bfm{p}_4$.  The complete
basis consists of four independent chiral amplitudes, each of them of
two possible isospin structures. For the moment we consider a parity
conserving theory, hence only two chiral amplitudes are not degenerate.
The Born amplitude is given by
\be
{\cal A}_{B}={ig^2\over s}{\cal A}^\lm\,.
\label{borna}
\ee
The collinear divergences in the hard part of the virtual corrections
and the corresponding {\it collinear} logarithms are known to factorize.
They are responsible, in particular, for the double logarithmic
contribution and depend only on the properties of the external on-shell
particles but not on the specific process
\cite{Mue1,Col,Sen1,CorTik,FreTay,APV,Sen2}.  This fact is especially
clear if a physical (Coulomb or axial) gauge is used for the
calculation. In this gauge the collinear divergences are present only in
the self energy insertions to the external particles
\cite{Sen1,FreTay,Sen2}.  Thus, for each fermion-antifermion pair of the
four-fermion amplitude the collinear logarithms are the same as for the
form factor ${\cal F}$ discussed in the previous section.  Let us denote
by $\tilde {\cal A}$ the amplitude with the collinear logarithms
factored out. For convenience we separate from $\tilde {\cal A}$ all the
corrections entering Eq.~(\ref{evolsolf}) so that
\be
{\cal A}={ig^2\over s}{\cal F}^2
\tilde{\cal A}\,.
\label{defa}
\ee
The resulting amplitude $\tilde {\cal A}$ contains the logarithms of the
{\it soft} nature corresponding to the soft divergences of the hard
region contribution\footnote{ Note that within the expansion by regions
the soft$\times$collinear double logarithmic divergence of the hard
region contribution is canceled against the ultraviolet divergence of the
collinear region {\it i.e.} there is no one-to-one correspondence
between the {\it soft logarithms} and the ultraviolet divergences of
the {\it soft region} contributions.},  and the renormalization group
logarithms.  It can be represented as a vector in the color/chiral basis
and satisfies the following evolution equation \cite{Sen2,Ste2,Bot}:
\be
{\partial \over \partial \ln{Q^2}}\tilde {\cal A}=
{\chi}(\al(Q^2))\tilde {\cal A} \,,
\label{evoleqa}
\ee
where $\chi(\al)$ is the matrix of the soft anomalous dimensions.  Note
that we do not include in Eq.~(\ref{evoleqa}) the pure renormalization
group logarithms which can be absorbed by fixing the renormalization scale
of $g$ in the Born amplitude~(\ref{borna}) to be $Q$.  The solution of
Eq.~(\ref{evoleqa}) is given by the path-ordered exponent
\be
\tilde {\cal A}=
{\rm P}\!\exp{\left[\int_{M^2}^{Q^2}
{\dd x\over x}\chi(\al(x))\right]}{\cal A}_{0}(\al(M^2))\,,
\label{evolsola}
\ee
where $\tilde {\cal A}_{0}(\al)$ determines the initial conditions for
the evolution equation at $Q=M$. The matrix of the soft anomalous
dimensions is determined by the coefficients of the single pole of the
hard region contribution to the exponent~(\ref{evolsola}).


In one loop the elements of the matrix $\chi(\al)$ do not depend on
chirality and read \cite{KPS}
\bea
\chi^{(1)}_{\lm \lm} &=&
-2C_A\left(\ln\left({x_+}\right)+i\pi\right)+
4\left(C_F-{T_F\over N}\right)\ln\left({x_+\over x_-}\right)\,, \nn \\
\chi^{(1)}_{\lm d} &=&4{C_FT_F\over N}\ln\left({x_+\over x_-}\right)\,, \nn \\
\chi^{(1)}_{d \lm} &=& 4\ln\left({x_+\over x_-}\right)\,,   \nn\\
\chi^{(1)}_{d d} &=& 0\,.
\label{chi}
\eea
In terms of the functions introduced above the one-loop correction reads
\bea
{\cal A}^{(1)}&=&{ig^2\over s}
\left[\left(\gm^{(1)}{\cal L}^2(Q^2)+
\left(2\xi^{(1)}+2\zeta^{(1)}+\chi^{(1)}_{\lm \lm}\right)
{\cal L}(Q^2)+2F_0^{(1)}\right){\cal A}^\lm\right.
\nn\\
&+&\left.\chi^{(1)}_{\lm d}
{\cal L}(Q^2){\cal A}^d+\tilde {\cal A}_0^{(1)}\right]\,,
\label{1loopa}
\eea
where
\be
\tilde {\cal A}_0^{(1)}=
\tilde A^{(1)}_0{}^\lm_{LL}{\cal A}^\lm_{LL}+
\tilde A^{(1)}_0{}^\lm_{LR}{\cal A}^\lm_{LR}+\ldots\,.
\label{deca}
\ee
For the present two-loop analysis of the annihilation cross section only
the real part of the coefficients $\tilde A_0^{(1)}$ is needed
(see {\it e.g.} \cite{KMPS}),
\bea
{\rm Re}\left[\tilde A^{(1)}_0{}^\lm_{LL}\right]&=&
\left(C_F-{T_F\over N}\right)f(x_+,x_-)
+{C_A}\left({85\over 9}+{\pi^2}\right)-{20\over 9}T_Fn_f-{8\over 9}T_Fn_s\,,
\nn \\
{\rm Re}\left[\tilde A^{(1)}_0{}^\lm_{LR}\right]&=&
-\left(C_F-{T_F\over N}-{C_A\over 2}\right)f(x_-,x_+)
+{C_A}\left({85\over 9}+{\pi^2}\right)-{20\over 9}T_Fn_f-{8\over 9}T_Fn_s\,,
\nn \\
{\rm Re}\left[\tilde A^{(1)}_0{}^d_{LL}\right]&=&
{C_FT_F\over N}f(x_+,x_-)\,,
\nn \\
{\rm Re}\left[\tilde A^{(1)}_0{}^d_{LR}\right]
&=&-{C_FT_F\over N}f(x_-,x_+)\,,
\label{rea0}
\eea
where
\be
f(x_+,x_-)={2\over x_+}\ln x_-+{x_--x_+\over x_+^2}\ln^2x_-\,.
\label{fpm}
\ee
The two-loop matrix $\chi^{(2)}$ can be extracted from the result for
the single pole part of the hard contribution to the four-quark
amplitude \cite{AGOT,Glo,FreBer}.  For vanishing beta-function it reads
\be
\chi^{(2)}|_{\bt_0=0}={\gm^{(2)}\over\gm^{(1)}}\chi^{(1)}\,,
\label{2loopchi}
\ee
After averaging over the isospin Eq.~(\ref{2loopchi}) agrees with the
result of Ref.~\cite{SteTej}.  An alternative approach to compute the
soft anomalous dimension matrix is based on studying the renormalization
properties of the product of light-like Wilson lines (see, {\em e.g.}
\cite{KOS}).  This approach has been used in Ref.~\cite{ADS} for the
calculation of the soft anomalous dimension matrix in two loops. The
result of Ref.~\cite{ADS} is in agreement with Eq.~(\ref{2loopchi}).

The part of the matrix $\chi^{(2)}$ proportional to the beta-function
originates from the running of the coupling constant in the one-loop
nonlogarithmic term.  We find that Eq.~(\ref{rea0}) gets contributions
only from the hard region so that the effective renormalization scale of
$\al$ there is $Q$.  Hence
\bea
\left.\chi^{(2)}_{\lm \lm}{\cal A}^\lm\right|_{\bt_0}
&=&-\bt_0\left(\tilde A^{(1)}_0{}^\lm_{LL}{\cal A}^\lm_{LL}+
\tilde A^{(1)}_0{}^{\lm}_{LR}{\cal A}^\lm_{LR}+\tilde A^{(1)}_0{}^d_{LL}{\cal A}^d_{LL}+
\tilde A^{(1)}_0{}^{d}_{LR}{\cal A}^d_{LR}\right)\,,
\label{2loopchibt}
\eea
{\it i.e.} in this order the soft anomalous dimension matrix does depend
on the chirality.  The matrix $\chi^{(1,2)}$ as well as the coefficients
$\tilde A^{(1)}_0{}^i_{IJ}$ are determined by the hard region
contribution and are  insensitive to the details of the gauge boson
mass generation.

Note that all the nontrivial soft region contributions in the expansion
of the multiscale two-loop integrals in the Sudakov limit are
power-suppressed \cite{SmiRak,Smi1,Smi2}. The unsuppressed soft region
contributions originate from one-scale self-energy insertions into an
external on-shell fermion or into a collinear gluon propagator.  On the
other hand one power of the logarithm in the double logarithmic
contribution is of soft nature.  Thus the two-loop soft logarithms in
$\tilde{\cal A}$ can be related to the collinear divergences though the
latter factorize and do not contribute to the soft anomalous dimension
matrix themselves.  This, in particular, explains the presence of the
$\gamma^{(2)}$ contribution to $\chi^{(2)}$.

The two-loop correction to the four-fermion amplitude is obtained by the
direct generalization of the form factor analysis with the only
complication related to the matrix structure of Eq.~(\ref{evolsola})
\bea
&&{\cal A}^{(2)}={ig^2(Q^2)\over s}\left\{{1\over 2}{(\gm^{(1)})}^2
{\cal L}^4{\cal A}^\lm+\left[\left(2\zeta^{(1)}+\chi^{(1)}_{\lm \lm}
-{1\over 3}\bt_0\right){\cal A}^\lm+\chi^{(1)}_{\lm d}{\cal A}^d\right]
{\gm^{(1)}}{\cal L}^3\right.
\nn\\
&&
+\left[\left(\gm^{(2)}+\left(2{\zeta^{(1)}}-\bt_0\right)
\zeta^{(1)}+2F_0^{(1)}\gm^{(1)}+{1\over 2}
\left(\left(4\zeta^{(1)}-\bt_0\right)\chi^{(1)}_{\lm \lm}
+{\chi^{(1)}_{\lm \lm}}^2+\chi^{(1)}_{\lm d}\chi^{(1)}_{d \lm}
\right)\right){\cal A}^\lm\right.
\nn\\
&&
\left. +{1\over 2}\left(\left(4\zeta^{(1)}-\bt_0\right)
\chi^{(1)}_{\lm d}+\chi^{(1)}_{\lm \lm}\chi^{(1)}_{\lm d}\right)
{\cal A}^d+\gm^{(1)}\tilde {\cal A}_0^{(1)}\right]{\cal L}^2
+\left[\left(2\zeta^{(2)}+2\xi^{(2)}+2F_0^{(1)}\left(2\zeta^{(1)}
\right.\right.\right.
\nn\\
&&
\left.\left.\left.+\chi^{(1)}_{\lm \lm}\right)+\chi^{(2)}_{\lm \lm}
\right){\cal A}^\lm + \left(2F_0^{(1)}\chi^{(1)}_{\lm d}
+\chi^{(2)}_{\lm d}\right){\cal A}^d
+\left(2\zeta^{(1)}+\chi^{(1)}\right)\tilde {\cal A}_0^{(1)}
\right]{\cal L}+{\cal O}({\cal L}^0)\bigg\}\,,
\label{nnnlla}
\eea
where we used the fact that $\xi^{(1)}=\chi^{(1)}_{dd}=0$ and  the matrix
structure of the product $\chi^{(1)}\tilde {\cal A}_0^{(1)}$ is implied.

Let us again discuss the standard model inspired example considered in
the previous section. With the result for the amplitudes it is
straightforward to compute the one- and two-loop corrections to the
total cross section of the four-fermion annihilation process.  For the
annihilation process $f'\bar f'\to f\bar f$ one has to make the
analytical continuation of the above result to the Minkowskian region of
negative $Q^2=-s$ according to the $s+i0$ prescription.  The above
approximation is formally not valid for the small angle region $\theta
<M/\sqrt{s}$, which, however, gives only a power-suppressed contribution
to the total cross section.  For the $SU(2)_L$ model we obtain
\bea
\sigma^{(2)}&=&\left[{9\over 2}{\cal L}^4(s)-{449\over 6}{\cal L}^3(s)
+\left({4855\over 18}+{37\over 3}\pi^2\right){\cal L}^2(s)
+\left({48049\over 216}-{1679\over 18}\pi^2-122\zeta(3)\right.\right.
\nn\\
&&
+15\sqrt{3}\pi
+26\sqrt{3}{\rm Cl}_2\left({\pi\over 3}\right)\bigg)
{\cal L}(s)\bigg]\sigma_B
\nn\\
&&
\approx\left(4.50\,{\cal L}^4(s)-74.83\,{\cal L}^3(s)+391.45\,{\cal L}^2(s)
-717.49\,{\cal L}(s)\right)\sigma_B\,,
\label{sigsu2s}
\eea
and
\bea
\sigma^{(2)}&=&\left[{9\over 2}{\cal L}^4(s)-{125\over 6}{\cal L}^3(s)
-\left({799\over 9}-{37\over 3}\pi^2\right){\cal L}^2(s)
+\left({38005\over 216}-{383\over 18}\pi^2-122\zeta(3)\right.\right.
\nn\\
&&+15\sqrt{3}\pi
+26\sqrt{3}{\rm Cl}_2\left({\pi\over 3}\right)\bigg)
{\cal L}(s)\bigg]\sigma_B
\nn\\
&&
\approx\left(4.50\,{\cal L}^4(s)-20.83\,{\cal L}^3(s)+32.95\,{\cal L}^2(s)
-53.38\,{\cal L}(s)\right)\sigma_B\,,
\label{sigsu2o}
\eea
for the initial and final state fermions of the same or opposite
eigenvalue of the isospin operator $t^3$, respectively.  Here $\sigma_B$
is the Born cross section with the $\overline{MS}$ coupling constants
renormalized at the scale $\sqrt{s}$ and ${\cal L}(s)=\ln(s/M^2)$.

\section{Two-loop  electroweak  logarithms}
\label{secew}
In Born approximation, the amplitude for the neutral current process
$f'\bar f'\rightarrow f\bar f$ reads
\be
{\cal A}_{B}={ig^2\over s}\sum_{I,J=L,R}\left(T^3_{f'}T^3_{f}+
t^2_W{Y_{f'}Y_{f}\over 4}\right){\cal A}^{f'f}_{IJ}\,,
\label{aborn}
\ee
where
\be
{\cal A}^{f'f}_{IJ}=\bar f_I'\gm_\mu f_I'
\bar f_J\gm_\mu f_J \,,
\label{defaff}
\ee
$t_W\equiv\tan{\theta_W}$ with $\theta_W$ being the Weinberg angle and
$T_f$ $(Y_f)$ is the isospin (hypercharge) of the fermion which depends
on the fermion chirality.  The one-loop electroweak correction to the
amplitude is well known (see \cite{BHM} for the most general result).
The calculation of the two-loop electroweak corrections even in the high
energy limit is a challenging theoretical problem at the limit of
available computational techniques.  It is complicated in particular by
the presence of the mass gap and mixing in the gauge sector.  In
Sect.~\ref{secsub} we develop the approach of Ref.~\cite{FKPS} and
reduce the analysis of the dominant two-loop logarithmic electroweak
corrections to a problem with a single mass parameter which has been
solved in the previous section.  In Sect.~\ref{secvir} we present the
explicit expression for the two-loop virtual linear-logarithmic
electroweak correction to the four-fermion amplitudes. In
Sect.~\ref{mdif} we complete the analysis of Refs.~\cite{KPS,KMPS} by
computing the leading effects of the $W-Z$ mass splitting through NNLL
approximation.  In Sect.~\ref{secrea} we introduce infrared safe
semi-inclusive cross sections and in Sect.~\ref{secnum} we give numerical
results for the two-loop corrections to the cross sections and various
asymmetries.

\subsection{Separating  QED infrared logarithms}
\label{secsub}
The main difference between the analysis of the electroweak standard
model with the spontaneously broken $SU_L(2)\times U(1)$ gauge group and
the treatment of the pure $SU_L(2)$ case considered above is the
presence of the massless photon which results in infrared divergences of
fully exclusive cross sections.  We regularize these divergences by
giving the photon a small mass $\lm$.  The dependence of the virtual
corrections on $\lm$ in the limit $\lm^2\ll M^2\ll Q^2$ is governed by
the QED infrared evolution equation and is given by the factor
\bea
&&\hspace*{-10pt}{\cal U}=U_0(\al_e(Q^2))
\nn\\
&&\times
\exp{\left\{{\alpha_e(\lm^2)\over 4\pi}\left[-\left(
Q_f^2+Q_{f'}^2-{\alpha_e\over \pi}\left({76\over 27}N_g
\left(Q_f^2+Q_{f'}^2\right)+{16\over 9}N_g\ln\left({x_+\over x_-}\right)
Q_{f'}Q_f\right)\right)\right.\right.}
\nn\\
&&\hspace*{-10pt}\times
\ln^2\left({Q^2\over \lm^2}\right)+\left[\left(3+{\alpha_e\over \pi}
\left(\left(-{40\over 9}+{16\over 27}\pi^2\right)N_g+
\left({3\over 8}-{\pi^2\over 2}+6\zeta(3)\right)Q_f^2\right)
\right)Q_f^2+\left(f\leftrightarrow f'\right)\right.
\nn\\
&&
\hspace*{-10pt}\left.\left.\left.
+\left(4-{\alpha_e\over \pi}{160\over 27}N_g\right)
\ln\left({x_+\over x_-}\right)Q_{f'}Q_f\right]\ln\left({Q^2\over\lm^2}
\right)-{\alpha_e\over \pi}{8\over 27}N_g\left(Q_f^2+Q_{f'}^2\right)
\ln^3\left({Q^2\over\lm^2}\right)\right]\right\},
\label{QED}
\eea
where $\al_e$ is the $\overline{MS}$ QED coupling constant and $Q_f$ is
the electric charge of the fermion.  The NNLL approximation for ${\cal
  U}$ can be found in \cite{KMPS} and to derive the two-loop linear
logarithm in the exponent~(\ref{QED}) we use the following expressions
for the QED parameters:
\bea
\zeta^{(2)}_e+\xi^{(2)}_e&=&\left[-{272\over 9}N_g+
\left({3\over 2}-2\pi^2+24\zeta(3)\right)Q_f^2\right]Q_f^2\,,
\nn\\
\chi^{(2)}_e|_{\bt_0=0}&=&
-{640\over 27}\ln\left({x_+\over x_-}\right)N_gQ_{f'}Q_f\,.
\label{QEDandim}
\eea
The expressions for $\zeta^{(2)}_e$, $\xi^{(2)}_e$ and $\chi^{(2)}_e$
can be obtained by substituting $n_f\to 8N_g/3$ into the general
formulae.  The coefficient $U_0(\al_e(Q^2))$ in Eq.~(\ref{QED}) is a
two-component vector in the chiral basis.  We have a freedom in
definition of this quantity because it does not depend on $\lm$ and can
be absorbed into the electroweak part of the corrections.  It is
convenient to normalize the QED factor to ${\cal U}|_{Q=\lm}=1$ so that
$U^{(1)}_0=0$.  Note that the structure of Eq.~(\ref{QED}) differs from
the one of the solution of the hard evolution
equations~(\ref{evolsolf},~\ref{evolsola}) because $\al_e$ in the
preexponential factor $U_0(\al_e)$ is renormalized at $Q$ so that all
the dependence on $\lm$ is contained in the exponent. By contrast, in
$F_0(\al)$ and $\tilde A_0(\al)$ the coupling constant is renormalized
at $M$ so the that all the dependence on $Q$ is contained in the
exponent.  This is why there is no $\chi^{(2)}_e|_{\bt_0}$ contribution
to the exponent~(\ref{QED}) and the $N_gQ_f^2$ part of the linear
logarithm coefficient there differs from the one of
$\zeta^{(2)}_e+\xi^{(2)}_e$.

Our goal now is to separate the above infrared divergent QED
contribution from the total two-loop corrections to get the pure
electroweak logarithms $\ln(Q^2/M_{W,Z}^2)$.  Within the evolution
equation approach \cite{KMPS} it has been found that the electroweak and
QED logarithms up to the NNLL approximation can be disentangled by means
of the following two-step procedure:
\begin{itemize}
\item[(i)] the corrections are evaluated using the fields of unbroken
  symmetry phase with all the gauge bosons of the same mass $M\approx
  M_{Z,W}$, {\it i.e.} without mass gap;
\item[(ii)] the QED contribution (\ref{QED}) with $\lm=M$ is factorized
  leaving the pure electroweak logarithms.
\end{itemize}
This reduces the calculation of the two-loop electroweak logarithms up
to the quadratic term to a problem with a single mass parameter.  Then
the effect of the $Z-W$ boson mass splitting can systematically be taken
into account within an expansion around the equal mass approximation
\cite{FKPS}.  In general the above two-step procedure is not valid in
the N$^3$LL approximation which is sensitive to fine details of the
gauge boson mass generation.  For the exact calculation of the
coefficient of the two-loop linear-logarithmic term one has to use the
true mass eigenstates of the standard model.  The evaluation of the
corrections in this case becomes a very complicated multiscale problem.
The analysis, however, is drastically simplified in a model with a Higgs
boson of zero hypercharge.  In this model the mixing is absent and the
above two-step procedure can be applied to disentangle all the two-loop
logarithms of the $SU_L(2)$ gauge boson mass from the infrared
logarithms associated with the massless hypercharge gauge boson (see the
discussion below).  In the standard model the mixing of the gauge bosons
results in a linear-logarithmic contribution which is not accounted for
in this approximation.  It is, however, suppressed by the small factor
$\sin^2\theta_W\equiv s_W^2\approx 0.2$.  Therefore, the approximation
gives an estimate of the coefficient in front of the linear electroweak
logarithm with $20\%$ accuracy.  As we will see, a $20\%$ error in the
coefficient in front of the two-loop linear electroweak logarithm leads
to an uncertainty comparable to the nonlogarithmic contribution and is
practically negligible.  If we also neglect the difference between $M_H$
and $M_{Z,W}$, the calculation involves a single mass parameter at every
step and the results of the previous sections can directly be applied to
the isospin $SU(2)_L$ gauge group with the coupling $g$ and the
hypercharge $U(1)$ gauge group with the coupling $t_Wg$.

The procedure of separating the QED contribution can naturally be
formulated in terms of the functions parameterizing the solution of the
hard evolution equation for the four-fermion amplitude.  The functions
$\gm(\al)$, $\zeta(\al)$, and $\chi(\al)$ are mass-independent.
Therefore, these functions parameterizing the electroweak logarithms can
be obtained by subtracting the QED quantities from the result for the
unbroken symmetry phase to all orders in the coupling constants ({\it
  cf.} Eq.~(\ref{hardmat})) without referring to the simplified model.
To the order of interest they can be found in \cite{KMPS} or easily
derived from the results of the previous sections.  For example, after
the subtraction, the $\bt_0=0$ part of the matrix $\chi^{(2)}$ for
$I,~J=L$ takes the form
\bea
\left.\chi^{(2)}_{\lm \lm}\right|_{\bt_0=0} &=&
\left(-{20\over 9}N_g+
{130\over 9}-{2\pi^2\over 3}\right)
\left[-4\left(\ln\left({x_+}\right)+i\pi\right)+
2\ln\left({x_+\over x_-}\right)\right]
\nn \\
&&+\left[-\left({100\over 27}N_g+{4\over 9}\right)
t_W^4Y_{f'}Y_f+{640\over 27}
N_gs_W^4Q_{f'}Q_f\right]\ln\left({x_+\over x_-}
\right)\,,
\nn \\
\left.\chi^{(2)}_{\lm d}\right|_{\bt_0=0}
&=&\left(-{5\over 3}N_g+
{65\over 6}-{\pi^2\over 2}\right)\ln\left({x_+\over x_-}\right)
\,,
\nn \\
\left.\chi^{(2)}_{d \lm}\right|_{\bt_0=0} &=& \left(-{80\over 9}N_g+
{520\over 9}-{8\pi^2\over 3}\right)\ln\left({x_+\over x_-}\right)
\,,
\nn \\
\left.\chi^{(2)}_{d d}\right|_{\bt_0=0}
 &=& \left[-\left({100\over 27}N_g+{4\over 9}\right)
t_W^4Y_{f'}Y_f+{640\over 27}N_gs_W^4Q_{f'}Q_f\right]
\ln\left({x_+\over x_-}\right)\,.
\label{subtrchina}
\eea
For $I$ or/and $J=R$ it  is reduced to
\be
\left.\chi^{(2)}\right|_{\bt_0=0}=
\left[-\left({100\over 27}N_g+{4\over 9}\right)t_W^4Y_{f'}Y_f
+{640\over 27}N_gs_W^4Q_{f'}Q_f\right]\ln\left({x_+\over x_-}\right)\,.
\label{subtrchia}
\ee
Note that no QED subtraction is necessary for $\chi^{(2)}|_{\bt_0}$
because of the specific normalization of the QED factor ${\cal U}$.
Thus $\xi^{(2)}$ is the only two-loop coefficient at the order of
interest which is sensitive to fine details of the gauge boson mass
generation. We evaluate it approximately by using the above simplified
model with the Higgs boson of zero hypercharge and of the mass
$M_H=M_{W,Z}$.  In this model the interference diagrams including the
heavy $SU_L(2)$ and the light hypercharge $U(1)$ gauge bosons are
identical to the ones of the $U(1)\times U(1)$ model discussed in
Sect.~\ref{secgap} and the corresponding contribution to $\xi^{(2)}$
vanishes.  At the same time the pure non-Abelian contribution to
$\xi^{(2)}$ is given by Eq.~(\ref{2loopx}).  Collecting all the pieces
for the two-loop contribution to the scheme-independent sum
$\zeta(\al)+\xi(\al)$ in the simplified model we obtain
\bea
\zeta^{(2)}+\xi^{(2)}
&=&
\Bigg[-{34\over 3}N_g+{749\over 12}+{43\over 18}\pi^2
-{176\over 3}\zeta(3)+5\sqrt{3}\pi
+{26\over 3}\sqrt{3}{\rm Cl}_2\left({\pi\over 3}\right)
\nn\\
&&
\left.+\left(3-4\pi^2+48\zeta(3)\right)t_W^2{Y_f^2\over 4}
\right]T_f(T_f+1)\,,
\label{subtrzt}
\eea
where $T_f=0$ and $1/2$ for the right- and left-handed fermions,
respectively.

\subsection{Two-loop virtual corrections}
\label{secvir}
The expressions for the two-loop corrections to the electroweak
amplitudes is obtained by projecting the result of Sect.~\ref{seca} on a
relevant initial/final state with the proper assignment of
isospin/hypercharge.  For example, the projection of the
basis~(\ref{basis}) on the states corresponding to the neutral current
processes reads ${\cal A}_{IJ}^\lm\to T^3_fT^3_{f'}{\cal A}_{IJ}^{f'f}$,
${\cal A}_{IJ}^d\to {\cal A}_{IJ}^{f'f}$.  The only complication in
combinatorics is related to the fact that now we are having different
gauge groups for the fermions of different chirality.  To take into
account the light fermion contribution one has to replace $n_f\to 2N_g$
for the $SU(2)_L$ and $n_f\to 5N_g/3$ for the hypercharge $U(1)$ gauge
group, where $N_g=3$ stands for the number of generations. The result
for the linear-logarithmic corrections is given in the following
approximation:
\begin{itemize}
\item[(i)] the fermion coupling to the Higgs boson is neglected, {\it
    i.e.} all the fermions are considered to be massless;
\item[(ii)] the $W-Z$ mass splitting is neglected;
\item[(iii)] the approximation (\ref{subtrzt}) is used for the two-loop
  coefficient $\zeta^{(2)}+\xi^{(2)}$.
\end{itemize}
The accuracy of the approximation is discussed in Sect.~(\ref{secnum}).
The result for the $n$-loop correction to the amplitude~(\ref{aborn})
can be decomposed as
\bea
{A}^{(n)} &=&{\cal A}^{(n)}_{LL} + {\cal A}^{(n)}_{NLL} +
{\cal A}^{(n)}_{NNLL}+\ldots\, .
\eea
Explicit expressions for the corrections up to ${\cal A}^{(2)}_{NNLL}$
can be found in~\cite{KPS,KMPS}.  The two-loop N$^3$LL term is new.  For
convenience we split it in parts as follows:
\bea
{\cal A}^{(2)}_{N^3LL}&=&
\sum_{i=1}^6\Delta_i{\cal A}^{(2)}_{N^3LL}\,.
\label{decannnl}
\eea
The correction corresponding to the $\xi^{(2)}$, $\zeta^{(2)}$ and
$F_0^{(1)}\zeta^{(1)}$ terms of Eq.~(\ref{nnnlla}) is
\bea
&&a^2\Delta_1{\cal A}^{(2)}_{N^3LL} = {ig^2\over
  s}\sum_{I,J=L,R}\! \Bigg\{\!
\Bigg[\Bigg(-{34\over 3}N_g+{749\over 12}+{43\over 18}\pi^2
-{176\over 3}\zeta(3)+5\sqrt{3}\pi
\nn\\
&&
\left.\left.+{26\over 3}\sqrt{3}{\rm Cl}_2\left({\pi\over 3}\right)
+\left(3-4\pi^2+48\zeta(3)\right)t_W^2{Y_f^2\over 4}\right)
\times T_f(T_f+1)
+ (f\leftrightarrow f')\right]
\nn\\
&&
-\left({21\over 2}+{2\pi^2}\right)
\left[T_f(T_f+1)+t_W^2{Y_f^2\over 4}-s_W^2Q_f^2+
(f\leftrightarrow f')\right]
\nn \\
&&
\times\left[T_f(T_f+1)+t_W^2{Y_f^2\over 4}
+(f\leftrightarrow f')\right]\Bigg\}\left(T^3_{f'}T^3_f+
t^2_W{Y_{f'}Y_f\over 4}\right)l(Q^2)a{\cal A}^{f'f}_{IJ} \,,
\label{nnnl1}
\eea
where $a = g^2/16\pi^2$ and $l(Q^2)=a\ln\left(Q^2/M^2\right)$.  The
correction corresponding to the $\chi^{(2)}$ term of Eq.~(\ref{nnnlla})
for vanishing beta-function reads
\bea
&&a^2\,\Delta_2{\cal A}^{(2)}_{N^3LL} \,=\, {ig^2\over
  s}\sum_{I,J=L,R}\Bigg\{\left(-{20\over 9}N_g+
{130\over 9}-{2\pi^2\over 3}\right)
\Bigg[\Bigg(-4\left(\ln\left(x_+\right)+i\pi\right)
\nn\\
&&
\left.\left.+\ln\left({x_+\over x_-}\right)\left(2+
t_W^2Y_{f'}Y_f\right)\right)T^3_{f'}T^3_f
+{3\over 4}\ln\left({x_+\over x_-}\right)\delta_{IL}\delta_{JL}\right]
+\ln\left({x_+\over x_-}\right)\!\!\left[-
\left({100\over 27}N_g+{4\over 9}\right)
\right.
\nn\\
&&
\left.\times t_W^4Y_{f'}Y_f+{640\over 27}N_gs_W^4Q_{f'}Q_f
\right]\left(T^3_{f'}T^3_f+t^2_W{Y_{f'}Y_f\over 4}\right)
\Bigg\}\,l(Q^2)aA^{f'f}_{IJ}\,,
\label{nnnl2}
\eea
while the $\bt_0$ part of $\chi^{(2)}$ term gives
\bea
&&a^2\,\Delta_3{\cal A}^{(2)}_{N^3LL} \,=\, {ig^2\over s}
\sum_{I,J=L,R}\left\{\left[\left({32\over 9}N_g-7\right)T^3_{f'}T^3_f+
\left({20\over 9}N_g+{1\over 6}\right)t^4_W{Y_{f'}Y_f\over 4}\right]
t_W^2{Y_{f'}Y_f\over 4}
\right.
\nn \\
&&
\times\bigg[f(x_+,x_-)\left(\delta_{IR}\delta_{JR}
+\delta_{IL}\delta_{JL}\right)-
f(x_-,x_+)\left(\delta_{IR}\delta_{JL}
+\delta_{IL}\delta_{JR}\right)\bigg]
\nn \\
&&-\left[
\left({80\over 27}N_g^2-{46\over 3}N_g-{86\over 27}\right)
T^3_{f'}T^3_f+
\left({2000\over 243}N_g^2+{130\over 81}N_g+{2\over 27}\right)
t_W^6{Y_{f'}Y_f\over 4}\right]+\left({4\over 3}N_g-{43\over 6}\right)
\nn \\
&&
\left.\times\left[
\left({1\over 2}f(x_+,x_-)+{170\over 9}+2\pi^2
\right)T^3_{f'}T^3_f
+{3\over 16}f(x_+,x_-)\delta_{IL}\delta_{JL}\right]\right\}\,l(Q^2)a
{\cal A}^{f'f}_{IJ}\,,
\label{nnnl3}
\eea
where $\delta_{IJ}=1$ for $I=J$ and zero otherwise.  Note that our
analysis implies the $\overline{MS}$ coupling constants renormalized at
the scale $M$ in the one-loop result with the exception of the one power
of the coupling constants originating from the Born
amplitude~(\ref{aborn}) renormalized at the scale $Q$.  The correction
corresponding to the $\zeta^{(1)}{\rm Re}\left[\tilde
  {A}^{(1)}_0\right]$ term of Eq.~(\ref{nnnlla}) is
\bea
&&a^2\,\Delta_4{\cal A}^{(2)}_{N^3LL} \,=\, {ig^2\over
  s}\sum_{I,J=L,R}{3}\left[T_f(T_f+1)+t_W^2{Y_f^2\over 4}-s_W^2Q_f^2+
(f\leftrightarrow f')\right]\left\{t_W^2{Y_{f'}Y_f\over 4}\right.
\nn\\
&&\times\left(2T^3_{f'}T^3_f+t^2_W{Y_{f'}Y_f\over 4}\right)
\bigg[f(x_+,x_-)
\left(\delta_{IR}\delta_{JR}+\delta_{IL}\delta_{JL}\right)-
f(x_-,x_+)\left(\delta_{IR}\delta_{JL}
+\delta_{IL}\delta_{JR}\right)\bigg]
\nn \\
&&-\left[
\left({20\over 9}N_g+{4\over 9}\right)T^3_{f'}T^3_f+
\left({100\over 27}N_g+{4\over 9}\right)t_W^4{Y_{f'}Y_f\over 4}\right]
+\left({1\over 2}f(x_+,x_-)+{170\over 9}+2\pi^2
\right)T^3_{f'}T^3_f
\nn \\
&&
+{3\over 16}f(x_+,x_-)\delta_{IL}\delta_{JL}\bigg\}\,l(Q^2)a
{\cal A}^{f'f}_{IJ}\,.
\label{nnnl4}
\eea
The correction corresponding to the $\chi^{(1)}{\rm Re}\left[\tilde
  {A}^{(1)}_0\right]$ term of Eq.~(\ref{nnnlla}) is
\bea
&&a^2\,\Delta_5{\cal A}^{(2)}_{N^3LL} \,=\, {ig^2\over  s}
\sum_{I,J=L,R}\Bigg\{\Bigg[\left(-8\left(\ln\left(x_+\right)+i\pi\right)
+\ln\left({x_+\over x_-}\right)\left(4+
t_W^2{Y_{f'}Y_f}\right)\right)T^3_{f'}T^3_f
\nn \\
&&
\left.
+{3\over 2}\ln\left({x_+\over x_-}\right)
\delta_{IL}\delta_{JL}
+\ln\left({x_+\over x_-}\right)\left(t_W^2Y_{f'}Y_f-
4s_W^2Q_{f'}Q_f\right)\left(2T^3_{f'}T^3_f+
t^2_W{Y_{f'}Y_f\over 4}\right)\right]t_W^2{Y_{f'}Y_f\over 4}
\nn \\
&&
\times\bigg[f(x_+,x_-)
\left(\delta_{IR}\delta_{JR}+\delta_{IL}\delta_{JL}\right)-
f(x_-,x_+)\left(\delta_{IR}\delta_{JL}
+\delta_{IL}\delta_{JR}\right)\bigg]
\nn \\
&&
+\left[\left({1\over 2}f(x_+,x_-)-{20\over 9}N_g+{166\over 9}
+2\pi^2\right)\left(-4\left(\ln\left(x_+\right)+i\pi\right)
+2\ln\left({x_+\over x_-}\right)\right)
\right.
\nn \\
&&
\left.+\ln\left({x_+\over x_-}\right)
\left(-\left({100\over 27}N_g+{4\over 9}\right)
t_W^4{Y_{f'}Y_f}+{3\over 4}f(x_+,x_-)\right)
\right]T^3_{f'}T^3_f
\nn \\
&&
+\left({3\over 8}f(x_+,x_-)-{5\over 3}N_g+{83\over 6}
+{3\over 2}\pi^2\right)\ln\left({x_+\over x_-}\right)
\delta_{IL}\delta_{JL}
\nn \\
&&+\left[\left({1\over 2}f(x_+,x_-)-{20\over 9}N_g+{166\over 9}
+2\pi^2\right)T^3_{f'}T^3_f-\left({100\over 27}N_g+{4\over 9}\right)
t_W^4{Y_{f'}Y_f\over 4}\right.
\nn \\
&&
\left.\left.+{3\over 16}f(x_+,x_-)
\delta_{IL}\delta_{JL}\right]\ln\left({x_+\over x_-}\right)
\left[t_W^2Y_{f'}Y_f-4s_W^2Q_{f'}Q_f\right]\right\}\,l(Q^2)a
{\cal A}^{f'f}_{IJ}\,.
\label{nnnl5}
\eea
The correction corresponding to the $\chi^{(1)}F_0^{(1)}$ term of
Eq.~(\ref{nnnlla}) reads
\bea
&&a^2\,\Delta_6{\cal A}^{(2)}_{N^3LL} \,=\, {ig^2\over s}
\sum_{I,J=L,R}-\Bigg\{\Bigg[\left(-4\left(\ln\left(x_+\right)+i\pi\right)
+\ln\left({x_+\over x_-}\right)\left(2+
t_W^2Y_{f'}Y_f\right)\right) T^3_{f'}T^3_f
\nn\\
&&
\left.
+{3\over 4}\ln\left({x_+\over x_-}\right)\delta_{IL}\delta_{JL}\right]
+\ln\left({x_+\over x_-}\right)\left(t_W^2Y_{f'}Y_f-
4s_W^2Q_{f'}Q_f\right)\left(T^3_{f'}T^3_f+t^2_W{Y_{f'}Y_f\over 4}\right)
\Bigg\}
\nn\\
&&\times\left({7\over 2}+{2\over 3}\pi^2\right)\left[T_f(T_f+1)
+t_W^2{Y_f^2\over 4}+(f\leftrightarrow f')\right]l(Q^2)a
{\cal A}^{f'f}_{IJ}\,.
\label{nnnl6}
\eea
The above expressions are derived for Euclidean positive $Q^2$.  For the
annihilation process one has to make the analytical continuation to the
Minkowskian region of negative $Q^2=-(s+i0)$ and it is natural to
normalize the QED factor to ${\cal U}|_{s=M^2}=1$ at the Minkowskian
point $s=M^2$ rather than at $Q^2=M^2$.  If we use this normalization
condition then ${\rm Re}\left[U^{(1)}_0\right]=-\pi^2(Q_f^2+Q_{f'}^2)$
and no QED subtraction is necessary for (the real part of) the amplitude
at $s=M^2$.

\subsection{The effect of $W-Z$ mass splitting}
\label{mdif}
The NLL and NNLL result for the electroweak corrections has been
obtained in Refs.~\cite{KPS,KMPS} neglecting the $Z-W$ boson mass
splitting which is suppressed by a small factor
$\delta_M=1-{M_W^2/M_Z^2}\approx 0.2$. The effect of the mass splitting
in the NLL and NNLL terms can, however, be comparable to the N$^3$LL
contribution and should be taken into account at this accuracy level.
In the NLL and NNLL approximation the mass splitting can easily be taken
into account through the modification of the one-loop parameters
$\xi^{(1)}$, $F_0^{(1)}$, and $\tilde {\cal A}_0^{(1)}$.  Let us take
$M_W$ as the argument of the logarithms.  Then the mass splitting
corrections originate from the $Z$ boson contribution. From the explicit
result for the one-loop diagrams in the leading order in $s_W^2$
and\footnote{ We do not express $\delta_M$ through $s_w^2$ to emphasize
  the origin of the corrections.}  $\delta_M\propto s_W^2$ we obtain
\bea
&&\delta\xi^{(1)}={2\over 3}T_f(T_f+1)\delta_M\,,
\nn\\
&&\delta F_0^{(1)}=-T_f(T_f+1)\delta_M\,,
\nn\\
&&\delta \tilde {\cal A}_0^{(1)}=-4\ln\left({x_+\over x_-}\right)
T^3_{f'}T^3_f\left(T^3_{f'}T^3_f+ t^2_W{Y_{f'}Y_f\over
4}\right)\delta_M{\cal A}^{f'f}_{LL}\,. \label{delm} \eea By using
Eq.~(\ref{delm}) it is straightforward to get the leading effect of
the mass splitting in one and two loops through the NNLL
approximation.  The corrections to the amplitudes read \bea
a\left.{\cal A}^{(1)}\right|_{\delta_M}&=&{ig^2\over
s}\sum_{I,J=L,R} \left[\left[T_f(T_f+1)+(f\leftrightarrow
f')\right]\left(-a+{2\over 3}
l(s)\right)-4T^3_{f'}T^3_f\ln\left({x_+\over x_-}\right)a\right]
\nn \\
&& \times\left(T^3_{f'}T^3_f+t^2_W{Y_{f'}Y_f\over
4}\right)\delta_M{\cal A}^{f'f}_{IJ}\,,
\nn\\
a^2\left.{\cal A}^{(2)}\right|_{\delta_M}&=&{ig^2\over
s}\sum_{I,J=L,R}\Bigg\{\left[ \left[T_f(T_f+1)+(f\leftrightarrow
f')\right]\left(-{2\over 3}L(s)l(s) +3l^2(s)\right)\right.
\nn\\
&& \left.+4T^3_{f'}T^3_f\ln\left({x_+\over x_-}\right)l^2(s)\right]
\left[T_f(T_f+1)+t_W^2{Y_f^2\over 4}-s_W^2Q_f^2+ (f\leftrightarrow
f')\right]
\nn \\
&& \times\left(T^3_{f'}T^3_f+t^2_W{Y_{f'}Y_f\over 4}\right)
+\left[\left(-4(\ln\left({x_+}\right)+i\pi) +\ln\left({x_+\over
x_-}\right)\left(2+ t_W^2Y_{f'}Y_f \right)\right)T^3_{f'}T^3_f
\right.
\nn\\
&&+{3\over 4}\ln\left({x_+\over x_-}\right)\delta_{IL}\delta_{JL}
+\ln\left({x_+\over x_-}\right)\left(t_W^2Y_{f'}Y_f-4s_W^2Q_{f'}Q_f
\right)\nn\\
&& \times\left(T^3_{f'}T^3_f+ t^2_W{Y_{f'}Y_f\over
4}\right)\Bigg]\left[T_f(T_f+1)+(f\leftrightarrow f')\right]{2\over
3}l^2(s)\Bigg\} \delta_M{\cal A}^{f'f}_{IJ}\,, \eea
which completes the result of Refs.~\cite{KPS,KMPS}.  Note that the
exponentiation of the mass splitting contribution due to
$\delta\xi^{(1)}$ has been observed first by the explicit calculation in
NLL approximation \cite{Poz}.

\subsection{Semi-inclusive cross sections}
\label{secrea}
To get the infrared safe result one has to take into account the real
photon emission in an inclusive way.  In practice, the massive gauge
bosons are supposed to be detected as separate particles.  Thus it is of
little physical sense to treat the hard photons with energies far larger
than $M_{Z,W}$ separately because of gauge symmetry restoration.  In
particular, the radiation of the hard real photons is not of the Poisson
type because of its non-Abelian $SU(2)_L$ component.  Therefore, we
restrict the analysis to semi-inclusive cross sections with the real
emission only of photons with energies far smaller than $M_{Z,W}$, which is
of pure QED nature.  To derive the result for such a cross section one
has to add to the expressions given above the QED corrections due to the
real photon emission and the {\it pure} QED virtual corrections which
are determined for $m_f\ll \lm\ll M$, where $m_f$ is a light fermion
mass, by Eqs.~(\ref{QED}).  However, in standard QED applications a
nonzero mass $m_f$ is kept for an on-shell fermion which regularizes the
collinear divergences. To derive the QED factor for $\lm$ far less than
the fermion mass $\lm\ll m_f\ll M$ one has to change the kernel of the
infrared evolution equation and match the new solution to
Eq.~(\ref{QED}) at the point $\lm =m_f$. For several light flavors of
significantly different masses the matching is necessary when $\lm$
crosses the value of each fermion mass.  The sum of the virtual and real
QED corrections to the cross section gives a factor which depends on
$s$, the fermion masses, and the experimental cuts, but not on $M_{Z,W}$.
The detailed analysis of the QED corrections goes beyond the scope of
the present paper. In the single flavor case the second order QED
corrections including the soft real emission are known beyond the
logarithmic approximation (see \cite{Pen} and references therein).

\subsection{Numerical estimates}
\label{secnum}
With the expressions for the chiral amplitudes at hand, we can compute
the total two-loop logarithmic virtual corrections to the basic
observables for the neutral current four-fermion processes.  Let us
consider the total cross sections of the quark-antiquark/$\mu^+\mu^-$
production in $e^+e^-$ annihilation.  In one and two loops the
logarithmic corrections read
\bea
R(e^+e^-\to Q\bar Q)&=&1-1.66\,L(s)+5.60\,l(s)-8.39\,a
\nn \\
&&
{}+1.93\,L^2(s)-11.28\,L(s)\,l(s)+33.79\,l^2(s)-60.87\,l(s)\,a \,,
\nn \\
R(e^+e^-\to q\bar q)&=&1-2.18\,L(s)+20.94\,l(s)-35.07\,a
\nn \\
&&
{}+2.79\,L^2(s)-51.98\,L(s)\,l(s)+321.20\,l^2(s)-757.35\,l(s)\,a\,,
\nn\\
R(e^+e^-\to
\mu^+\mu^-)&=&1-1.39\,L(s)+10.35\,l(s)-21.26\,a
\nn \\
&&
{}+1.42\,L^2(s)-20.33\,L(s)\,l(s)+112.57\,l^2(s)-312.90\,l(s)\,a\,,
\label{finres}
\eea
where $L(s)=a\ln^2\left(s/M^2\right)$, $Q=u,c,t$, $q=d,s,b$,
$R(e^+e^-\to \mu^+\mu^-)=\sigma/\sigma_B(e^+e^-\rightarrow \mu^+\mu^-)$
and so on. The $\overline{MS}$ couplings in the Born cross section are
renormalized at $\sqrt{s}$.  Numerically, we have $L(s)=0.07$ $(0.11)$
and $l(s)=0.014$ $(0.017)$ for $\sqrt{s}=1$~TeV and $2$~TeV,
respectively.  Here $M=M_W$ has been chosen for the infrared cutoff and
$a = 2.69\cdot 10^{-3}$, $s_W^2=0.231$ for the $\overline{MS}$ couplings
renormalized at the gauge boson mass.  The complete one-loop corrections
are known exactly and we have included the dominant one-loop terms in
Eqs.~(\ref{finres})--(\ref{finreslrtil}) to demonstrate the structure of
the expansion rather than for precise numerical estimates though the
above expressions approximate the exact one-loop result with $1\%$
accuracy in the TeV region.  In Eq.~(\ref{finres}) we included the
leading correction in the $W-Z$ mass difference through NNLL
approximation. The coefficient of the linear logarithm is computed in
the approximation of Sect.~(\ref{secsub}).  In the case of a
quark-antiquark final state the strong interaction could also produce
logarithmically growing terms. For massless quarks the complete ${\cal
  O}(\al_s^2)$ corrections including the bremsstrahlung effects can be
found in \cite{KraLam}.  Note that Eqs.~(\ref{finres}) are symmetric
under exchange of the initial and final state fermions and, therefore,
also applicable to the Drell-Yan processes at hadron colliders.  In this
case the two-loop QCD corrections are given in \cite{MMN}.  For the
total cross sections of the four-quark electroweak processes we obtain
\bea
R(Q'\bar Q'\to Q\bar Q)&=&1-2.07\,L(s)+19.03\,l(s)-32.63\,a
\nn\\
&&
{}+2.67\,L^2(s)-46.64\,L(s)\,l(s)+278.94\,l^2(s)-666.05\,l(s)\,a\,,
\nn\\
R(Q\bar Q\to q\bar q)&=&1-2.56\,L(s)+8.49\,l(s)-11.94\,a
\nn \\
&&
{}+3.53\,L^2(s)-20.39\,L(s)\,l(s)+65.20\,l^2(s)-91.92\,l(s)\,a \,,
\nn \\
R(q'\bar q'\to q\bar q)&=&1-2.87\,L(s)+25.63\,l(s)-38.89\,a\nn \\
&&
{}+4.20\,L^2(s)-71.87\,L(s)\,l(s)+423.61\,l^2(s)-919.35\,l(s)\,a\,.
\label{finresqq}
\eea
Our results can easily be generalized to $f\bar f \to f\bar f$ processes
with identical quarks by including the $t$-channel contribution.

For $e^+e^-$ annihilation we also give a numerical estimate of
corrections to the cross section asymmetries.  In the case of the
forward-backward asymmetry $A^{FB}$ (the difference of the cross section
integrated over forward and backward hemispheres with respect to the
electron beam direction divided by the total cross section) we get
\bea
R^{FB}(e^+e^-\to Q\bar Q)&=&1-0.09\,L(s)-1.22\,l(s)+1.77\,a
\nn \\
&&
{}+0.12\,L^2(s)+0.59\,L(s)\,l(s)-1.65\,l^2(s)
+3.39\,l(s)\,a\,,
\nn \\
R^{FB}(e^+e^-\to q\bar q)&=&1-0.14\,L(s)+7.17\,l(s)-10.84\,a
\nn \\
&&
{}+0.02\,L^2(s)-1.32\,L(s)\,l(s)-33.07\,l^2(s)
+93.44\,l(s)\,a\,,
\nn\\
R^{FB}(e^+e^-\to\mu^+\mu^-)&=&1-0.04\,L(s)+5.50\,l(s)-14.43\,a
\nn \\
&&
{}+0.27\,L^2(s)-6.43\,L(s)\,l(s)+22.91\,l^2(s)
-33.11\,l(s)\,a\,,
\label{finresfb}
\eea
where $R^{FB}=A^{FB}/A^{FB}_B$.  For the left-right asymmetry $A^{LR}$
(the difference of the cross sections of the production of left- and
right-handed particles divided by the total cross section) we obtain in
the same notations
\bea R^{LR}(e^+e^-\to Q\bar Q)&=&1-4.48\,L(s)+17.51\,l(s)-13.16\,a
\nn \\
&& {}-1.16\,L^2(s)+15.66\,L(s)\,l(s)-43.50\,l^2(s)
+44.05\,l(s)\,a\,,
\nn \\
R^{LR}(e^+e^-\to q\bar q)&=&1-1.12\,L(s)+12.05\,l(s)-16.44\,a
\nn \\
&&
{}-0.81\,L^2(s)+18.02\,L(s)\,l(s)-130.74\,l^2(s)
+278.71\,l(s)\,a\,,
\nn\\
R^{LR}(e^+e^-\to\mu^+\mu^-)&=&1-13.24\,L(s)+116.58\,l(s)-148.42\,a
\nn \\
&&
{}-0.79\,L^2(s)+23.68\,L(s)\,l(s)-155.46\,l^2(s)
-116.67\,l(s)\,a\,.
\nn \\
&&
\label{finreslr}
\eea

\begin{figure}
  \begin{center}
    \begin{tabular}{cc}
      \hspace*{-9mm}
      \epsfxsize=8.5cm
      \epsffile{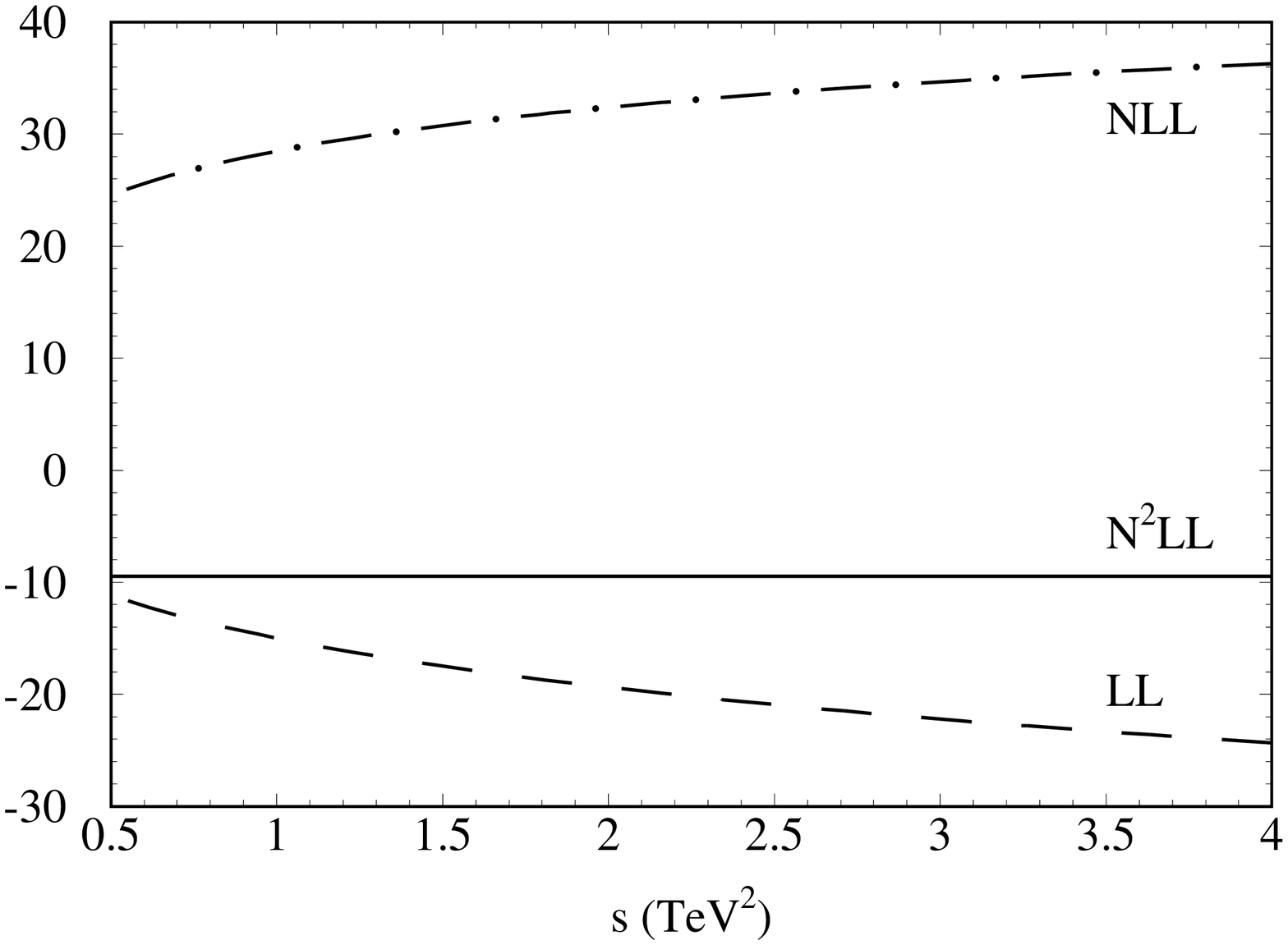}&
      \hspace*{-9mm}
      \epsfxsize=8.5cm
      \epsffile{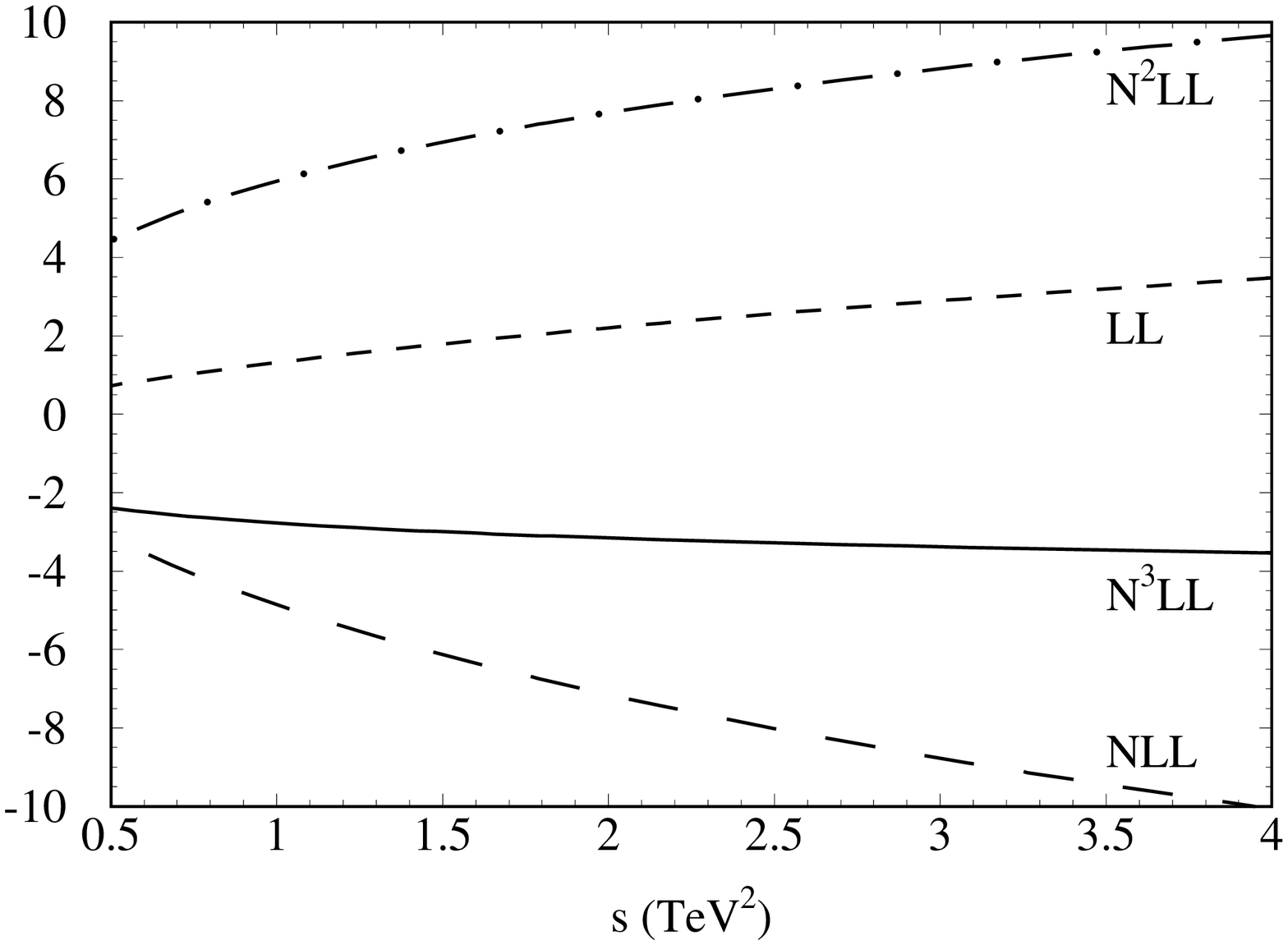}\\
      \hspace*{-9mm}
      (a)&\hspace*{-9mm}(b)
    \end{tabular}
  \end{center}
  \caption{\label{fig1} \small Separate logarithmic contributions
    to $R(e^+e^-\to q\bar q)$ in \% to the Born approximation: (a) the
    one-loop LL $(\ln^2(s/M^2)$, long-dashed line), NLL $(\ln^1(s/M^2)$,
    dot-dashed line) and N$^2$LL $(\ln^0(s/M^2)$, solid line) terms; (b)
    the two-loop LL $(\ln^4(s/M^2)$, short-dashed line), NLL
    $(\ln^3(s/M^2)$, long-dashed line), NNLL $(\ln^2(s/M^2)$, dot-dashed
    line) and N$^3$LL $(\ln^1(s/M^2)$, solid line) terms.}
\end{figure}

Finally, for the left-right asymmetry $\tilde A^{LR}$ (the difference of
the cross sections for the left- and right-handed initial state particles
divided by the total cross section) which differs from $A^{LR}$ for the
quark-antiquark final state we have
\bea
\tilde R^{LR}(e^+e^-\to Q\bar Q)&=&1-2.75\,L(s)+10.60\,l(s)-9.05\,a
\nn \\
&&
{}-0.91\,L^2(s)+11.16\,L(s)\,l(s)-33.49\,l^2(s)
+28.28\,l(s)\,a\,,
\nn \\
\tilde R^{LR}(e^+e^-\to q\bar q)&=&1-1.07\,L(s)+11.75\,l(s)-16.21\,a
\nn \\
&&
{}-0.77\,L^2(s)+17.06\,L(s)\,l(s)-125.18\,l^2(s)
+267.60\,l(s)\,a\,.
\nn \\
&&
\label{finreslrtil}
\eea

\begin{figure}
  \begin{center}
    \begin{tabular}{cc}
      \hspace*{-9mm}
      \epsfxsize=8.5cm
      \epsffile{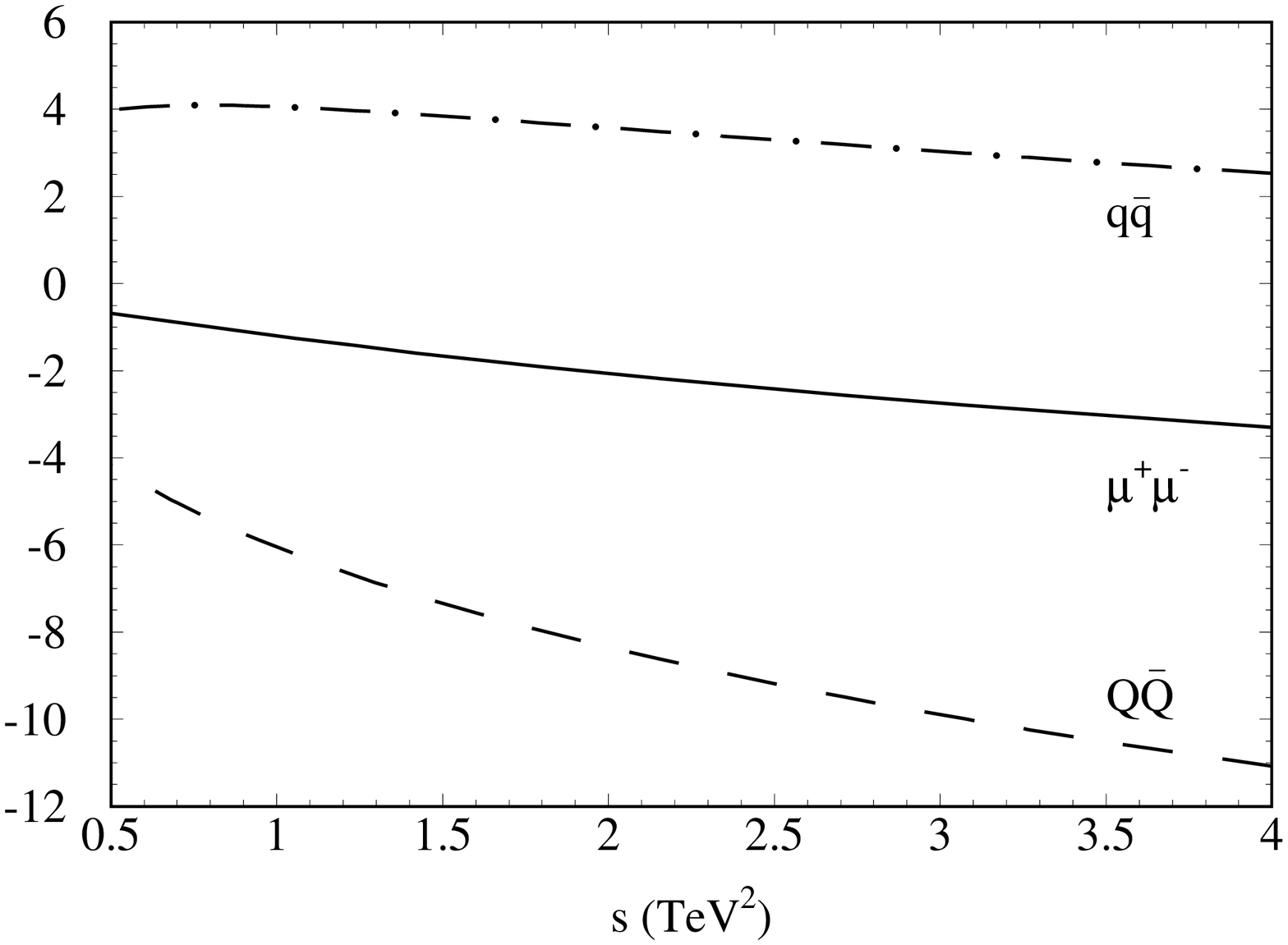}&
      \hspace*{-9mm}
      \epsfxsize=8.5cm
      \epsffile{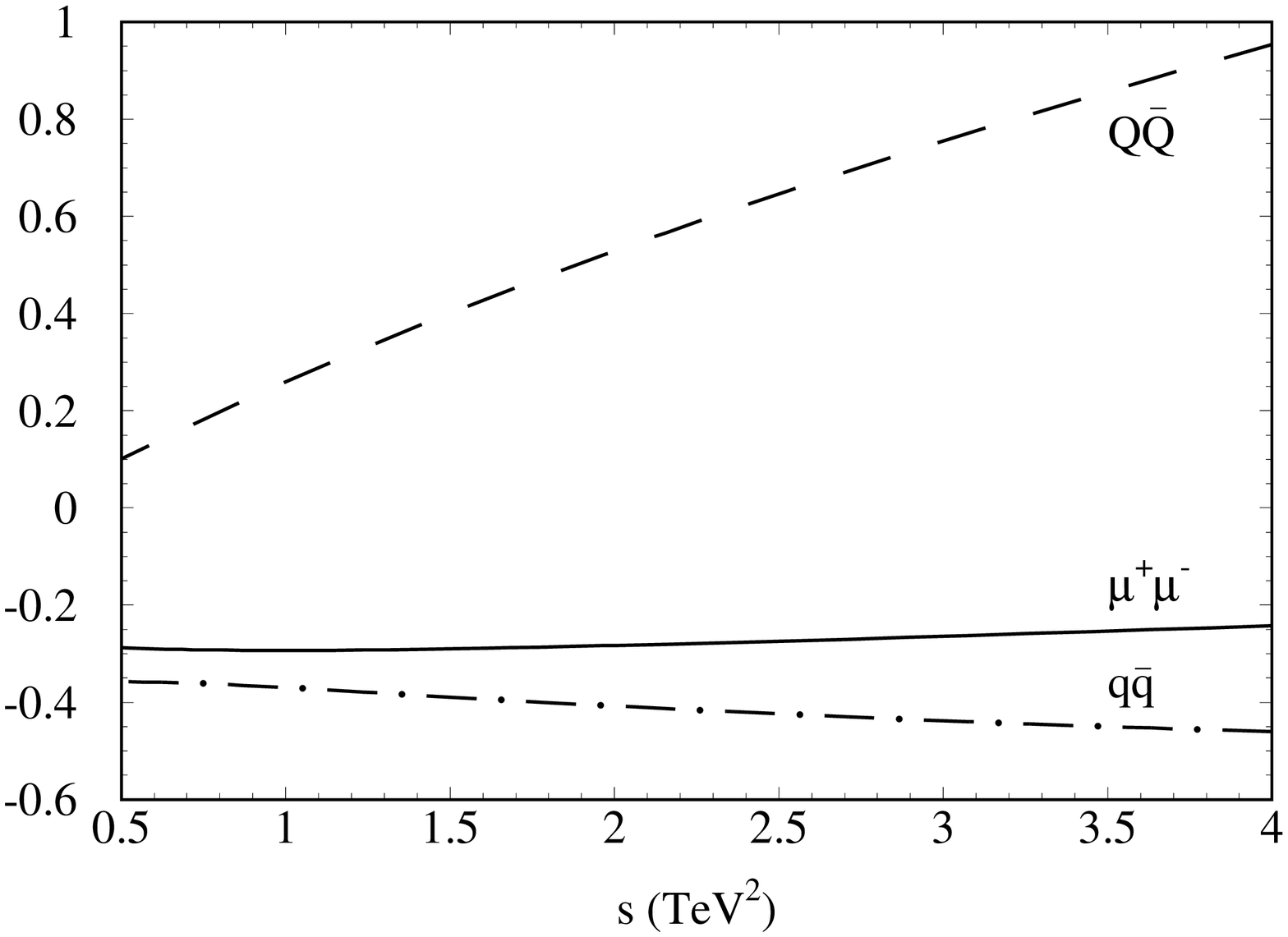}\\
      \hspace*{-9mm}
      (a)&\hspace*{-9mm}(b)
    \end{tabular}
  \end{center}
  \caption{\label{fig2} \small
    The total logarithmic corrections to $R(e^+e^-\to Q\bar Q)$ (dashed
    line), $R(e^+e^-\to q\bar q)$ (dot-dashed line) and $R(e^+e^-\to
    \mu^+\mu^-)$ (solid line) in \% to the Born approximation: (a) the
    one-loop correction up to N$^2$LL term; (b) the two-loop correction
    up to N$^3$LL term.}
\end{figure}

\begin{figure}
  \begin{center}
    \begin{tabular}{cc}
      \hspace*{-9mm}
      \epsfxsize=8.5cm
      \epsffile{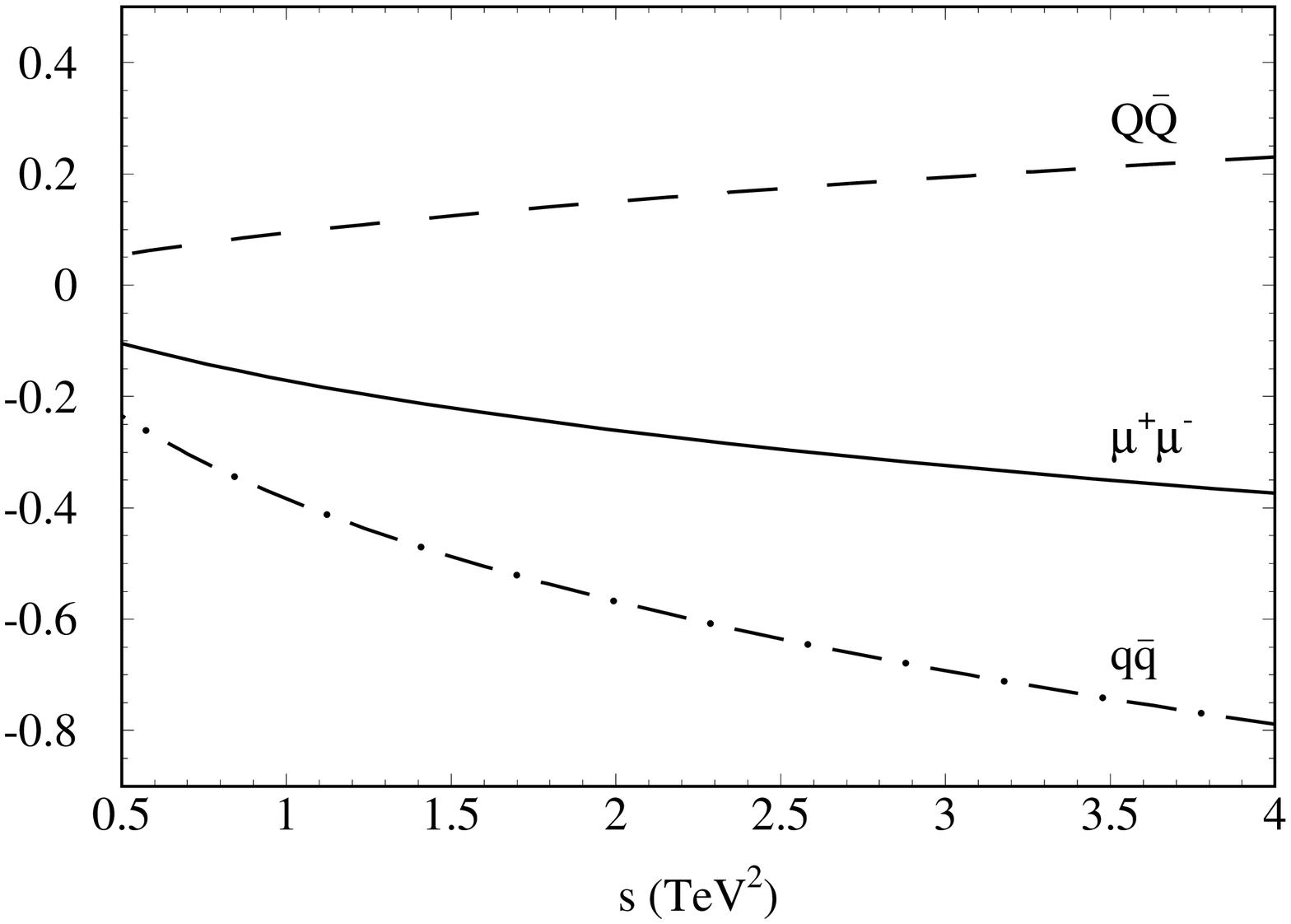}&
      \hspace*{-9mm}
      \epsfxsize=8.5cm
      \epsffile{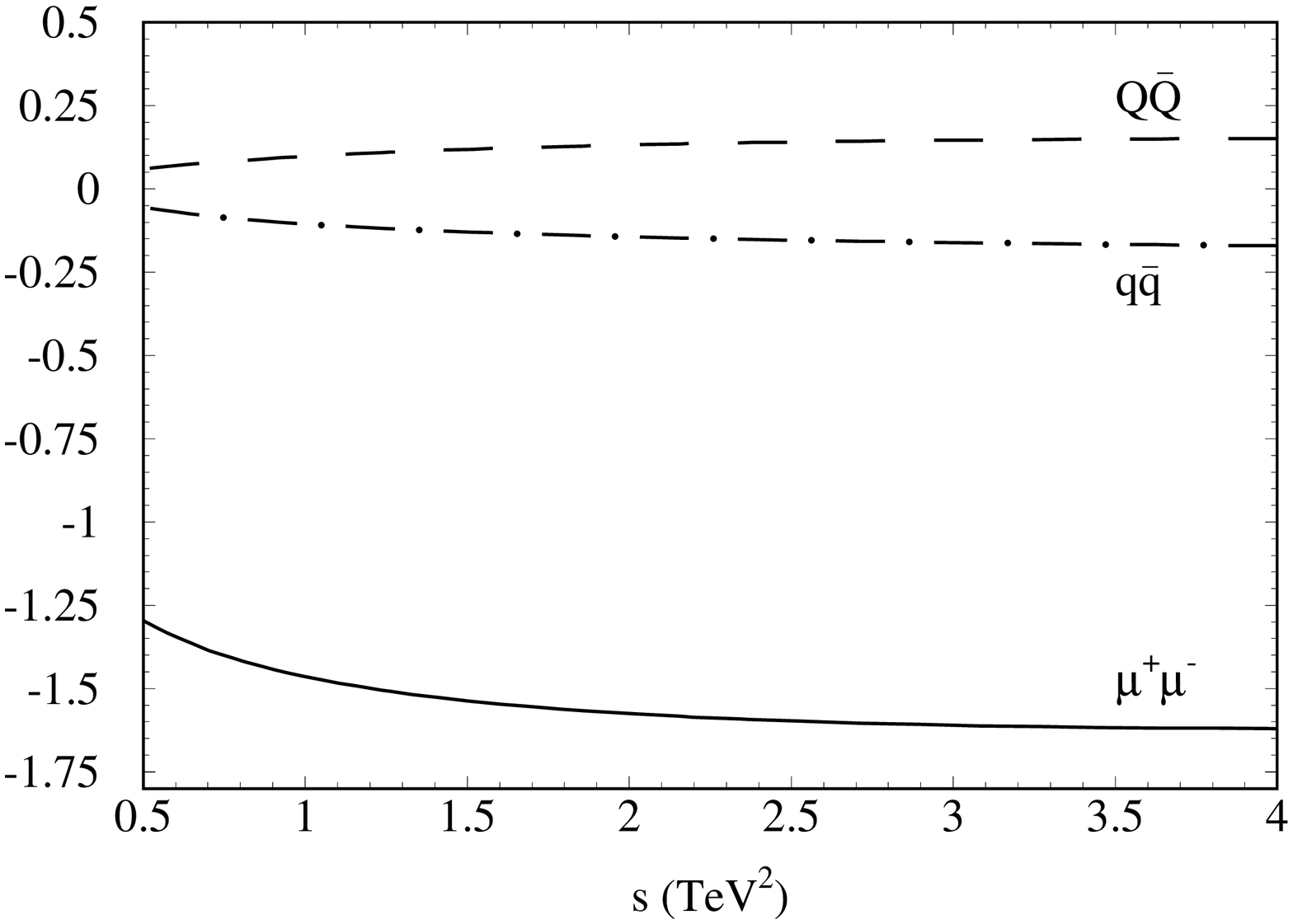}\\
      \hspace*{-9mm}
      (a)&\hspace*{-9mm}(b)
    \end{tabular}
  \end{center}
  \caption{\label{fig3} \small
    (a) The total logarithmic two-loop corrections to the
    forward-backward asymmetry $R^{FB}(e^+e^-\to Q\bar Q)$ (dashed
    line), $R^{FB}(e^+e^-\to q\bar q)$ (dot-dashed line) and
    $R^{FB}(e^+e^-\to \mu^+\mu^-)$ (solid line) in \% to the Born
    approximation. (b) The same as (a) but for the left-right asymmetry
    $\tilde R^{LR}$.}
\end{figure}

The numerical structure of the corrections in the case of $e^+e^-$
annihilation is shown in Figs.~\ref{fig1}-\ref{fig3}.  In
Fig.~\ref{fig1} the values of different logarithmic contributions to
$R(e^+e^-\to q\bar q)$ are plotted separately as functions of $s$.  In
the $1-2$~TeV region the two-loop LL, NLL, NNLL and N$^3$LL corrections
to the cross sections can be as large as $1-4\%$, $5-10\%$, $5-10\%$ and
$2-3\%$ respectively. The two-loop logarithmic terms have a
sign-alternating structure resulting in significant cancellations
typical for the Sudakov limit \cite{KMPS,FKM,FKPS}.  Although the
individual logarithmic contributions can be as large as $10\%$, their
sum does not exceed $1\%$ in absolute value at energies below 2~TeV
for all the cross sections (see Fig.~\ref{fig2}).  In the region of a
few TeV the corrections do not reach the double-logarithmic asymptotics.
The quartic, cubic and quadratic logarithms are comparable in magnitude.
Then the logarithmic expansion starts to converge.  Still, the
linear-logarithmic contribution must be included to reduce the
theoretical uncertainty below $1\%$. The two-loop logarithmic
corrections to the asymmetries also amount up to $1\%$ in absolute
value at energies below 2~TeV with the only exception of
$R^{LR}(e^+e^-\to\mu^+\mu^-)$ (see Fig.~\ref{fig3}).

Let us discuss the accuracy of our result.  On the basis of the explicit
evaluation of the light fermion/scalar \cite{FKM} and the Abelian
contribution \cite{FKPS} we estimate the uncalculated two-loop
nonlogarithmic term to be at a few permill level.  The power-suppressed
terms do not exceed a permill in magnitude for $\sqrt{s}>500$~GeV as
well \cite{FKM}.  By comparing our numerical estimates to the equal mass
approximation given in the Appendix we find that the leading effect of
the $W-Z$ mass splitting results in a variation of at most $5\%$ of the
coefficients of the two-loop cubic and quadratic logarithms. Thus the
expansion in the $W-Z$ mass difference converges well for these
coefficients and the leading correction term taken into account in our
evaluation is sufficient for a permill accuracy of the cross sections.
The main uncertainty of our result for the two-loop single logarithmic
contribution is due to our approximation for the $\xi^{(2)}$
coefficient. Neglecting the gauge boson mixing effects, which are
suppressed by a factor of $\sin^2{\theta_W}$, brings an error of $20\%$
to the coefficient of the two-loop single logarithm.  Neglecting the
difference between the Higgs and gauge boson masses does not lead to a
numerically important error because the scalar boson contribution is
relatively small.  By comparing the results for $M_H=M_{W,Z}$ and
$M_H\ll M_{W,Z}$ we estimate the corresponding uncertainty in the
coefficient of the two-loop single logarithm to be about $5\%$. The same
estimate is true for the uncertainty due to the top quark mass effect on
the $t\bar t$ virtual pair contribution. Hence for the production of
light fermions our formulae are supposed to approximate the exact
coefficients of the two-loop linear logarithms with approximately $20\%$
accuracy which results in a few permill uncertainty in the cross
sections.  By adding up the errors from different sources in quadrature
we find the total uncertainty of the cross section to be in a few
permill -- one percent range depending on the process.  Thus we get an
accurate estimate of the two-loop correction, which is sufficient for
practical applications to the future collider physics.  The only
essential deviation of the exact two-loop logarithmic contributions from
our result Eqs.~(\ref{finres}--\ref{finreslrtil}) for the production of
the third generation quarks is due to the large top quark Yukawa
coupling.  The corresponding corrections are known to NLL approximation
and can numerically be as important as the generic non-Yukawa ones
\cite{Bec2,DenPoz,Mel2,BecMir}.

\section{Summary}
\label{summ}
In the present paper we have derived the analytical result for the
two-loop logarithmic corrections to the vector form factor and the
four-fermion cross sections in the spontaneously broken $SU_L(2)$ model
by combining the explicit calculation and the evolution equation
approach.  We have completed the analysis of the dominant
logarithmically enhanced two-loop electroweak corrections to the basic
observables of the neutral current four-fermion processes at high
energy.  The two-loop linear-logarithmic contribution has been obtained
with an estimated accuracy of $20\%$. The $W-Z$ mass splitting effect
neglected in \cite{KPS,KMPS} has been taken into account through the
NNLL approximation.  Our result for the two-loop logarithmic corrections
along with the known exact one-loop expressions approximates the cross
sections with an accuracy of a few permill, which is sufficient for
practical applications to the future collider physics.  The general
approach for the calculation of the logarithmically enhanced corrections
developed in the present paper can also be applied to the analysis of
the gauge boson production and the supersymmetric extensions of the
standard model where only the NLL approximation is available so far
\cite{Bec3,Bec4}.

\vspace{4mm}

{\bf Acknowledgments}\\[3mm]

We thank the authors of Ref.~\cite{ADS} for pointing out a spurious term
in the previous version of Eq.~(41) caused by a misinterpretation of the
results of Refs.~\cite{Glo,FreBer}.  We would like to thank the authors
of Ref.~\cite{ADS} also for sending us their result prior to publication
and for a very helpful discussion.  We gratefully acknowledge a useful
communication with the authors of Refs.~\cite{Glo,SteTej}.  The work of
J.H.K and A.A.P.  was supported in part by BMBF Grant No.\ 05HT4VKA/3
and Sonderforschungsbereich Transregio 9.  The work of V.A.S. was
supported in part by the Russian Foundation for Basic Research through
project 05-02-17645 and DFG Mercator Grant No.  Ha 202/110-1.  The work
of B.J. was supported in part by Cusanuswerk,
Landesgraduiertenf\"orderung Baden-W\"urttemberg and the DFG
Graduiertenkolleg ``Hochenergiephysik und Teilchenastrophysik''.

\section*{Appendix}
The NNLL result for the cross sections and asymmetries in the
approximation $M_Z=M_W$ can be obtained from the expressions of
Ref.~\cite{KMPS}\footnote{Throughout Sect~4. of Ref.~\cite{KMPS} the
  terms with the factor $(aN_g+b)t_W^2$, where $a$ and $b$ stand for
  some constants, should be multiplied by an extra $t_W^2$, and the
  terms with the factor $N_gs_W^2$ should be multiplied by an extra
  $s_W^2$. This results in a small change of the numerical estimates.}
\bea
R(e^+e^-\to Q\bar Q)&=&1-1.66\,L(s)+5.31\,l(s)-8.36\,a
\nn \\
&&
{}+1.93\,L^2(s)-10.59\,L(s)l(s)+31.40\,l^2(s) \, ,
\nn \\
R(e^+e^-\to q\bar q)&=&1-2.18\,L(s)+20.58\,l(s)-34.02\,a
\nn \\
&&
{}+2.79\,L^2(s)-51.04\,L(s)l(s)+309.34\,l^2(s)\, ,
\nn\\
R(e^+e^-\to\mu^+\mu^-)&=&1-1.39\,L(s)+10.12\,l(s)-20.61\,a
\nn \\
&&
{}+1.42\,L^2(s)-19.81\,L(s)l(s)+107.03\,l^2(s)\, ,
\label{finresap}
\eea
\bea
R^{FB}(e^+e^-\to Q\bar Q)&=&1-0.09\,L(s)-1.23\,l(s)+1.47\,a
\nn \\
&&
{}+0.12\,L^2(s)+0.64\,L(s)l(s)-1.40\,l^2(s)\, ,
\nn \\
R^{FB}(e^+e^-\to q\bar q)&=&1-0.14\,L(s)+7.15\,l(s)-10.43\,a
\nn \\
&&
{}+0.02\,L^2(s)-1.31\,L(s)l(s)-33.46\,l^2(s)\, ,
\nn\\
R^{FB}(e^+e^-\to\mu^+\mu^-)&=&1-0.04\,L(s)+5.49\,l(s)-14.03\,a
\nn \\
&& {}+0.27\,L^2(s)-6.32\,L(s)l(s)+21.01\,l^2(s)\, ,
\label{finresfbap} \eea \bea R^{LR}(e^+e^-\to Q\bar
Q)&=&1-4.48\,L(s)+16.66\,l(s)-13.28\,a
\nn \\
&& {}-1.16\,L^2(s)+15.21\,L(s)l(s)-41.79\,l^2(s) \, ,
\nn \\
R^{LR}(e^+e^-\to q\bar q)&=&1-1.12\,L(s)+11.86\,l(s)-15.83\,a
\nn \\
&&
{}-0.81\,L^2(s)+17.74\,L(s)l(s)-127.05\,l^2(s)\, ,
\nn\\
R^{LR}(e^+e^-\to\mu^+\mu^-)&=&1-13.24\,L(s)+113.77\,l(s)-139.94\,a
\nn \\
&&
{}-0.79\,L^2(s)+23.34\,L(s)l(s)-155.36\,l^2(s)\, .
\label{finreslrap}
\eea
\bea
\tilde R^{LR}(e^+e^-\to Q\bar Q)&=&1-2.75\,L(s)+10.07\,l(s)-9.02\,a
\nn \\
&&
{}-0.91\,L^2(s)+10.80\,L(s)l(s)-32.10\,l^2(s) \, ,
\nn \\
\tilde R^{LR}(e^+e^-\to q\bar q)&=&1-1.07\,L(s)+11.56\,l(s)-15.60\,a
\nn \\
&&
{}-0.77\,L^2(s)+16.78\,L(s)l(s)-121.56\,l^2(s)\, .
\label{finreslrtilap}
\eea


\begin{thebibliography}{99}
\bibitem{Sud}    V.V. Sudakov, {Zh. Eksp. Teor. Fiz.} {30} (1956) 87.

\bibitem{Jac}    R. Jackiw, {Ann. Phys.} {48} (1968) 292; {51} (1969) 575.

\bibitem{Kur}    M. Kuroda, G. Moultaka, and D. Schildknecht, {Nucl. Phys.}
                 {B 350} (1991) 25.

\bibitem{DegSir} G. Degrassi and A. Sirlin, {Phys. Rev.} {D 46} (1992) 3104.

\bibitem{Bec}    M. Beccaria {\it et al.}, {Phys. Rev.} {D 58} (1998)
                 093014.

\bibitem{CiaCom} P. Ciafaloni and  D. Comelli,  {Phys. Lett.} {B 446}
                 (1999) 278.

\bibitem{KuhPen} J.H. K\"uhn and A.A. Penin, Report TTP99--28 and
                 hep-ph/9906545.

\bibitem{Fad}    V.S. Fadin, L.N. Lipatov, A.D. Martin, and M. Melles,
                 {Phys. Rev.} {D 61} (2000)  094002.

\bibitem{KPS}    J.H. K\"uhn, A.A. Penin, and V.A. Smirnov,
                 {Eur. Phys. J.}  {C 17} (2000) 97;
                 {Nucl. Phys. B (Proc. Suppl.)} {89} (2000) 94.

\bibitem{KMPS}   J.H. K\"uhn, S. Moch, A.A. Penin, and V.A. Smirnov,
                 {Nucl.Phys.} {B 616} (2001) 286, Erratum {\it  ibid.}
                 {B 648} (2003)  455.



\bibitem{Bec2}   M. Beccaria {\it et al.},  {Phys. Rev.} {D 61} (2000)
                 011301; {D 61} (2000)  073005.

\bibitem{BRV}    M. Beccaria, F.M. Renard, and C. Verzegnassi,
                 {Phys. Rev.} {D 63} (2001) 053013.

\bibitem{DenPoz} A. Denner and S. Pozzorini,
                 {Eur. Phys. J.}  {C 18} (2001) 461; C {21} (2001) 63.

\bibitem{HKK}    M. Hori, H. Kawamura, and J. Kodaira,
                 {Phys. Lett.} {B 491} (2000) 275.

\bibitem{BeeWet} W. Beenakker, A. Werthenbach {Nucl.Phys} {B 630} (2002) 3.

\bibitem{FKM}    B. Feucht, J.H. K\"uhn, and S. Moch, {Phys. Lett. }  B { 561}, 111 (2003).

\bibitem{DMP}    A. Denner, M. Melles, and  S. Pozzorini,  {Nucl. Phys.}
                 {B662} (2003) 299.

\bibitem{Bec3}   M. Beccaria, F.M. Renard, and C. Verzegnassi,
                 {Nucl. Phys.} {B 663} (2003) 394.

\bibitem{Bec4}   M. Beccaria {\it et al.},
                 {Int. J. Mod. Phys.} {A 18} (2003) 5069.

\bibitem{Poz}    S. Pozzorini, {Nucl. Phys.} {B692} (2004) 135.

\bibitem{FKPS}   B. Feucht, J.H. K\"uhn, A.A. Penin, and V.A. Smirnov,
                 Phys. Rev. Lett. {93} (2004) 101802.


\bibitem{BenSmi} M. Beneke and V.A. Smirnov, {Nucl. Phys.} {B 522} (1998) 321.

\bibitem{SmiRak} V.A. Smirnov and E.R. Rakhmetov, {Teor. Mat. Fiz.}  {120} (1999) 64.

\bibitem{Smi1}   V.A. Smirnov, {Phys. Lett. }  B { 465} (1999) 226.

\bibitem{Smi2}   V.A. Smirnov, {\it Applied Asymptotic Expansions in Momenta and Masses}
                 (Springer-Verlag, Berlin, Heidelberg, 2001).

\bibitem{Mue1}   A.H. Mueller {Phys. Rev.} {D 20} (1979) 2037.

\bibitem{Col}    J.C. Collins, {Phys. Rev.} {D 22} (1980) 1478;
                 in {\it Perturbative QCD}, ed. A.H. Mueller, 1989, p. 573.

\bibitem{Sen1}   A. Sen, {Phys. Rev.} {D 24} (1981) 3281.

\bibitem{JKPS} B. Jantzen, J.H. K\"uhn, A.A. Penin, and V.A. Smirnov,
                 {Phys. Rev.} {D 72} (2005) 051301(R).

\bibitem{Ste1}   G. Sterman {Phys. Rev.}  {D 17} (1978) 2773.

\bibitem{LibSte} S. Libby and G. Sterman {Phys. Rev.} {D 18} (1978) 3252.

\bibitem{Mue2}   A.H. Mueller, {Phys. Rep.} {73} (1981) 35.

\bibitem{Smi3}   V.A. Smirnov, {\it Evaluating Feynman Integrals}
                 (Springer-Verlag, Berlin, Heidelberg, 2004).

\bibitem{Smi4}   V.A. Smirnov, {Phys. Lett.} {B  547} (2002) 239.

\bibitem{KraLam} G. Kramer and B. Lampe,  {Z. Phys.} {C 34} (1987) 497,
                 Erratum {\it  ibid.} {C 42} (1989) 504.

\bibitem{MMN}    T. Matsuura, S.C. van der Marck, and W.L. van Neerven,
                 {Nucl. Phys.} {B 319} (1989) 570.


\bibitem{MagSte} L. Magnea and G. Sterman, {Phys. Rev.} {D 42} (1990) 4222.

\bibitem{KorRad} G.P. Korchemsky and A.V. Radyushkin,
                 {Nucl. Phys.} {B283} (1987) 342.

\bibitem{MVV}    S.Moch, J.A.M. Vermaseren, and A. Vogt,
                 {Nucl. Phys.} {B 688} (2004) 101.


\bibitem{Kin}    T. Kinoshita, {J. Math. Phys.} {3} (1962) 650.

\bibitem{LeeNau} T.D. Lee and M. Nauenberg,  {Phys. Rev.}
                 D {133} (1964) 1549.

\bibitem{CorTik} J.M. Cornwall and  G. Tiktopoulos, { Phys. Rev. Lett.}
                 {35} (1975) 338; {Phys. Rev.} {D 13} (1976) 3370.

\bibitem{FreTay} J. Frenkel and J.C. Taylor, {Nucl. Phys.}
                 {B 116} (1976) 185.

\bibitem{APV}    D. Amati, R. Petronzio, and G. Veneziano,
                 {Nucl. Phys.} {B 146} (1978) 29.

\bibitem{Sen2}   A. Sen, {Phys. Rev.} {D 28} (1983) 860.

\bibitem{Ste2}   G. Sterman, {Nucl. Phys.} {B 281} (1987) 310.

\bibitem{Bot}    J. Botts and G. Sterman, {Nucl. Phys.} {B 325} (1989) 62.

\bibitem{AGOT}   C. Anastasiou, E.W.N. Glover, C. Oleari, and
                 M.E. Tejeda-Yeomans,  {Nucl. Phys.} {B 601} (2001) 341.

\bibitem{Glo}    E.W.N. Glover, {JHEP} {0404}  (2004) 021.

\bibitem{FreBer} A. De Freitas and  Z. Bern, {JHEP} {0409} (2004) 039.

\bibitem{SteTej} G. Sterman and M.E. Tejeda-Yeomans,
                 {Phys. Lett.} {B 552} (2003) 48.

\bibitem{KOS}    N. Kidonakis, G. Oderda, and G. Sterman,
                {Nucl. Phys.} {B 531} (1998) 365.

\bibitem{ADS}    M. Aybat, L. Dixon,  and G. Sterman,  Report
                 No. YITP-SB-06-25, SLAC-PUB-11907, and  hep-ph/0606254.





\bibitem{BHM}    W. Beenakker, W. Hollik, and S.C. Van der Marck,
                 {Nucl. Phys.} {B 365} (1991) 24.


\bibitem{Pen}    A.A. Penin,   {Phys. Rev. Lett.} {95} (2005) 010408;
                 {Nucl. Phys.} {B 734},  185 (2006).




\bibitem{Mel2}   M. Melles, {Phys. Rev.} {D 64} (2001) 014011.

\bibitem{BecMir} M. Beccaria and E. Mirabella,  Phys. Rev. {D 72} (2005)
                 055004.

\end{thebibliography}
\end{document}